\documentclass[11pt,a4paper]{article}
\pdfoutput=1

\usepackage{jheppub}
\usepackage{color}
\usepackage[dvipsnames]{xcolor}
\usepackage{graphicx}
\usepackage{wrapfig}
\usepackage{verbatim}
\usepackage{amsmath}
\usepackage{amssymb}
\usepackage{url}
\usepackage{bbold}
\usepackage{xspace}
\usepackage{slashed}
\usepackage{multirow}
\usepackage{threeparttable}
\usepackage{paralist}
\usepackage{subcaption}
\usepackage{floatrow}
\usepackage{multirow}
\usepackage{soul}

\usepackage{tikz}
\usetikzlibrary{trees}
\usetikzlibrary{decorations.pathmorphing}
\usetikzlibrary{decorations.markings}
\usetikzlibrary{arrows}

\newcommand{\MeV}{\,\mathrm{MeV}}
\newcommand{\GeV}{\,\mathrm{GeV}}
\newcommand{\TeV}{\,\mathrm{TeV}}

\newcommand{\vev}[1]{\langle #1 \rangle}

\newcommand{\beq}{\begin{equation}}
\newcommand{\eeq}{\end{equation}}
\newcommand{\bea}{\begin{eqnarray}}
\newcommand{\eea}{\end{eqnarray}}

\DeclareRobustCommand{\Sec}[1]{Sec.~\ref{#1}}

\DeclareRobustCommand{\App}[1]{App.~\ref{#1}}

\DeclareRobustCommand{\Fig}[1]{Fig.~\ref{#1}}
\DeclareRobustCommand{\Figs}[2]{Figs.~\ref{#1} and \ref{#2}}
\DeclareRobustCommand{\Eq}[1]{Eq.~(\ref{#1})}
\DeclareRobustCommand{\Eqs}[2]{Eqs.~(\ref{#1}), (\ref{#2})}

\newcommand{\roll}{\Lambda_{\text{\tiny R}}}
\newcommand{\br}{\Lambda_{\text{\tiny B}}}
\newcommand{\tildebr}{\tilde{\Lambda}_{\text{\tiny B}}}

\newcommand{\thetam}{\theta_*}
\newcommand{\ellm}{\ell_*}
\newcommand{\Rs}{R_{\text{\tiny S}}}
\newcommand{\Rt}{R_{\text{\tiny T}}}
\newcommand{\DRt}{\Delta R_{\text{\tiny T}}}
\newcommand{\mroll}{\mu}
\newcommand{\km}{\,\mathrm{km}}
\newcommand{\Rwd}{R_{\text{\tiny WD}}}
\newcommand{\nwd}{n_{\text{\tiny WD}}}
\newcommand{\Rns}{R_{\text{\tiny NS}}}
\newcommand{\nns}{n_{\text{\tiny NS}}}
\newcommand{\mplanck}{M_{\text{\tiny P}}}
\newcommand{\barr}{\Delta V_{\text{top}}}
\newcommand{\qcd}{\Lambda_{\text{\tiny QCD}}}
\newcommand{\thetacp}{\theta_{\text{\tiny QCD}}}
\newcommand{\noqcd}{\Lambda_{\text{\tiny C}}}
\newcommand{\Rnsd}{R_{\tilde{\text{\tiny S}}}}
\newcommand{\mdark}{m_{\tilde{b}}}
\newcommand{\ndark}{n_{\tilde{b}}}
\newcommand{\tc}{\Lambda_{\text{\tiny TC}}}
\newcommand{\Os}{\Omega_{\text{\tiny S}}}
\newcommand{\Bs}{B_{\text{\tiny S}}}
\newcommand{\Ons}{\Omega_{\text{\tiny NS}}}
\newcommand{\Bns}{B_{\text{\tiny NS}}}
\newcommand{\Rtem}{R_{\text{\tiny T}}^{\text{\tiny EM}}}
\newcommand{\mN}{m_N}

\begin{document}

{\large
\flushright TUM-HEP-1348/21 \\ }

\title{Runaway Relaxion from Finite Density}

\author[a,b]{Reuven Balkin,}
\author[a]{Javi Serra,}
\author[a]{Konstantin Springmann,}
\author[a]{Stefan Stelzl,}
\author[a]{Andreas Weiler}

\affiliation[a]{Physik-Department, Technische Universit\"at M\"unchen, 85748 Garching, Germany}
\affiliation[b]{Physics Department, Technion -- Israel Institute of Technology, Haifa 3200003, Israel}

\date{\today}

\abstract{Finite density effects can destabilize the metastable vacua in relaxion models. Focusing on stars as nucleation seeds, we derive the conditions that lead to the formation and runaway of a relaxion bubble of a lower energy minimum than in vacuum. The resulting late-time phase transition in the universe allows us to set new constraints on the parameter space of relaxion models. We also find that similar instabilities can be triggered by the large electromagnetic fields around rotating neutron stars.}

\preprint{}
\maketitle

%%%%%%%%%%%%%%%%%%%%%%%%%%%%%%%%%%%%%%%%%%%
%%%%%%%%%%%%%%%%%%%%%%%%%%%%%%%%%%%%%%%%%%%
\newpage

\section{Introduction} \label{sec:intro}

Expectations based on effective field theory are seemingly failing to reveal the resolution of the naturalness problem of the electroweak scale. The absence of the long-sought signals of new physics at the LHC strongly motivates the investigation of alternatives that defy those expectations.
One possibility is to consider that the Higgs mass is in fact not fundamentally fixed but instead can take different values, depending on the local vacuum where the theory is realized.
The discovery that our vacuum is just one of many in a landscape would have profound implications for our understanding of particle physics, placing our universe on the same footing as our planet, solar system or galaxy.
The multiple-vacua hypothesis was put to use by Weinberg as a solution of the cosmological constant problem \cite{Weinberg:1987dv}, and in recent years it has gained attention as an explanation for a hierarchically small electroweak scale, see e.g.~\cite{Graham:2015cka,Arvanitaki:2016xds,Herraez:2016dxn,Geller:2018xvz,Cheung:2018xnu,Giudice:2019iwl,Kaloper:2019xfj,Strumia:2020bdy,Csaki:2020zqz,Arkani-Hamed:2020yna,Giudice:2021viw} as well as \cite{Agrawal:1997gf,Dvali:2003br,ArkaniHamed:2004fb,Dvali:2004tma,ArkaniHamed:2005yv} for earlier related proposals.
A paradigmatic example by now is the relaxion \cite{Graham:2015cka}, where the Higgs mass not only takes different values depending on the local minimum where the relaxion sits, but the structure of the landscape is such that a mechanism operating during cosmological evolution can be engineered to dynamically select the right vacuum.
While these types of scenarios come with some theoretical problems (vacuum selection, dependence on initial conditions, measure problem), more important is the fact that they give rise to distinct experimental consequences and signatures, thus to new ways of testing the origin of the electroweak scale.

In this work we wish to explore one of the novel phenomenological aspects associated with a landscape of vacua as that present in relaxion models. 
Irrespective of how the multiple vacua are populated or our vacuum is selected, it is well-known that when the scalar field parametrizing the landscape is in a thermal bath, the structure of its potential changes, allowing for transitions between different vacua.
This phenomenon can take place as well at finite density if the scalar is coupled to the background matter, as recently discussed in \cite{Hook:2019pbh,Balkin:2021zfd}.
This gives rise to the possibility that large and dense systems, such as stars, induce a classical (or quantum) transition from the metastable minimum, where the theory in vacuum resides, to a lower energy minimum.%
\footnote{Some of these results have been presented in several talks over the last two years \cite{KStalk:2020Jan,SStalk:2020Sep,RBtalk:2020Oct}.}
Interestingly, if certain conditions on the density profile and the evolution of the star are met, the corresponding scalar bubble can expand beyond the confines of the dense system, potentially leading to a phase transition where the universe transitions to the new, energetically preferred, ground state.
Since this type of late-time phase transitions, taking place at the dawn of star formation (redshifts $z \lesssim 20$), are constrained by cosmological observations, they lead to bounds on the properties of the landscape.

Both the conditions for a scalar bubble to escape from a star and the experimental constraints on the associated phase transition have been discussed in \cite{Balkin:2021zfd} in the context of a simple quartic potential with a tilt, i.e.~with two non-degenerate minima. In this work we extend them to the case of the relaxion potential, which we review in \Sec{sec:potential}.
Since by construction the relaxion potential depends on the QCD quark condensate (QCD-relaxion) or the Higgs VEV squared (non-QCD relaxion), its landscape of minima changes in a dense environment of SM matter.
We show in \Sec{sec:denserelaxion} that at sufficiently high densities and for large enough stars, the in-vacuo relaxion minimum can be destabilized and a bubble of a lower energy minimum can expand indefinitely.
This fact allows us to set new constraints on the relaxion mechanism, which we present in \Sec{sec:relaxionbounds}. The most relevant ones are found in the case of the non-QCD relaxion (\Sec{sec:relaxionnoqcd}), where the minima are typically very shallow and the small change in the Higgs VEV induced by a finite baryon density is enough to trigger the transition.
We also find that the large electric and magnetic fields generated by pulsars/magnetars can lead to analogous phase transitions for relaxions with large couplings to photons (\Sec{sec:relaxiontc}).
Finally, we show that if the vacuum instability is seeded by the largest stars in the universe (with radii thousand times that of the Sun), or if dark astrophysical objects exist (\Sec{sec:dark}), a change of minimum in the relaxion landscape can take place with the corresponding change in vacuum energy being very small. A priori, such a phase transition is phenomenologically viable and an interesting target for future exploration.

We summarize and present our conclusions in \Sec{sec:conclusions}.

%%%%%%%%%%%%%%%%%%%%%%%%%%%%%%%%%%%%%%%%%%
%%%%%%%%%%%%%%%%%%%%%%%%%%%%%%%%%%%%%%%%%%
\section{The relaxion potential} \label{sec:potential}

The relaxion potential \cite{Graham:2015cka} is characterized by a washboard-like shape where the amplitude of the wiggles depends on the relaxion field $\phi$ itself,
\beq
V(\phi) =  - \roll^4 \, \frac{\phi}{f} - \tildebr^4 \, F(\phi) \cos \frac{\phi}{f} \,. %+ \dots \,.
\label{eq:Vgeneric}
\eeq
$\roll$ and $\tildebr$ are the scales that control the size of the linear rolling and periodic back-reaction terms respectively, while $2\pi f$ parametrizes the field distance between adjacent minima.
The monotonically increasing function $F(\phi)$ is of the form 
\beq
F(\phi) = \left(\frac{\phi}{\phi_c}-1\right)^{k/2} \Theta(\phi-\phi_\text{c}) \,,
\label{eq:F}
\eeq
with $k=1,2$ and where $\phi_c$ is the field value where the periodic barriers turn on.
This is taken such that the change in the size of the wiggles after a $2\pi f$ period is small, i.e.~$\phi_c \gg f$, and therefore the landscape is densely populated over field ranges of order $\phi_c$. The case $k=0$ corresponds to Abbott's potential \cite{Abbott:1984qf}, where the size of the potential barriers is constant.
The non-trivial behavior for $k=1,2$ arises from the dependence of the periodic term on the Higgs VEV, $h$, which in turn is determined by the scalar field $\phi$.

Indeed, the Higgs potential, in particular the mass term, depends on the value of the relaxion,
\beq
V(h) = \tfrac{1}{2} (M^2 - g \phi M) h^2 + \tfrac{1}{4} \lambda h^4 \, ,
\label{eq:higgs}
\eeq
where $M$ is the cutoff and $g$ is a small coupling that breaks the shift-symmetry associated with $\phi$. Note that the periodic term in the relaxion potential, while breaking the continuous shift-symmetry, is still invariant under the discrete shift $\phi \to \phi + 2 \pi f n$, $n \in \mathbb{Z}$. The coupling to the Higgs, as well as the linear term in the potential, break it completely, thus we expect
\beq
\frac{\roll^4}{f} = c \, g M^3 \, ,
\label{eq:rollg}
\eeq
where $c$ is a parameter with the dimensions of an inverse coupling squared, therefore $c \sim 1/(4\pi)^2$ in a strongly coupled UV completion; in this work we simply take $c = 1$. Likewise, naive dimensional analysis yields $M \sim 4 \pi f$; here we keep $M$ and $f$ independent and require that $f > M/4\pi$ ($f \gg M$ would correspond to a weakly coupled UV completion).
As soon as the Higgs mass parameter turns negative, $h$ acquires a VEV, given by
\beq
h^2 = \frac{M^2}{\lambda} \left(\frac{\phi}{\phi_c} - 1\right) \, , \quad \phi_c \equiv M/g \, ,
\label{eq:higgsvev}
\eeq
where we have identified $\phi_c$ as given in \Eq{eq:F}. 

One must now specify how the amplitude of the relaxion periodic term depends on the Higgs VEV.
In \Sec{sec:relaxionbounds} we discuss specific realizations of the relaxion, where such a dependence is either linear (QCD-relaxion), $\tildebr^4 F(\phi) \equiv \qcd^4 h/v$ with $v \approx 246 \GeV$ and $\qcd$ the QCD quark condensate, or quadratic (non-QCD-relaxion), $\tildebr^4 F(\phi) \equiv \noqcd (h/v)^2$, where $\noqcd$ is the analogous of $\qcd$ for a new QCD-like confining dynamics.
These two cases therefore correspond to $k = 1, 2$ in \Eq{eq:F}, respectively. 
More complicated functions, beyond \Eq{eq:F}, arise in the presence of extra light scanning scalars \cite{Espinosa:2015eda}.
In any case, the change in the Higgs field between adjacent minima is 
\beq
\Delta h^2 = \frac{2\pi}{\lambda} \frac{\roll^4}{M^2} \, .
\label{eq:higgschange}
\eeq
The requirement that $\phi_c \gg f$, which is correlated with the fact that the rolling term is hierarchically smaller than the cutoff of the theory, $\roll \ll M$, see \Eq{eq:rollg}, ensures that the Higgs VEV varies slowly with every period of the potential.

Recall that in the relaxion mechanism, $\phi$ goes through a period of dynamical evolution, originally assumed to happen during a phase of cosmological inflation \cite{Graham:2015cka}, where it rolls towards the minima of the landscape, generically stopping at one of the first (see below for a characterization of the minima). The parameters of the potential are adjusted, in a technically natural fashion, such that the Higgs VEV is of the right size at the minimum where the relaxion stops its evolution, that is $h = v$.
Other proposals regarding the aforementioned time evolution of the relaxion have been put forward in e.g.~\cite{Hardy:2015laa,Hook:2016mqo,Fonseca:2018xzp} and \cite{Fonseca:2019lmc}.
In addition, already in \cite{Graham:2015cka} and subsequently in e.g.~\cite{Espinosa:2015eda,Nelson:2017cfv}, the potential itself was made to evolve during inflation, eventually leading to the relaxion resting in a minimum many periods beyond the first.
In this work we will not be concerned with the early cosmological dynamics of the relaxion. Instead, our analysis generally applies to whichever minimum the relaxion eventually stopped at, i.e.~to the minimum where the relaxion is found when structures in the universe, in particular stars, start to form.

To ease the analysis of the landscape of relaxion minima, it is useful to redefine the scalar field as
\beq
\phi \equiv \phi_\ell(\theta) = \left( 2\pi \ell + \theta \right) f \;\;\; \text{with} \;\;\; \ell \in \mathbb{N}\,, \;\; \theta \in [0,2\pi)\,,
\label{eq:field}
\eeq
where $\ell$ labels the period of the field. The local (metastable) minima of the potential are then denoted by $\phi_{\ellm}(\thetam)$, where the precise value of $\thetam$ depends on the period $\ellm$. Minima are found as soon as the ($\ellm$-dependent) effective back-reaction grows large enough,
\beq
\br^4 \equiv \tildebr^4 F(\phi_{\ellm}(\pi/2)) > \roll^4 \,.
\label{eq:rescale}
\eeq
In addition, we can conveniently choose to shift the origin of field space such that the minima start with $\ellm = 1$, $\phi_\ell \to \phi_\ell + 2 \pi f (\bar{\ell}-1)$ with $\bar{\ell} = (\phi_c/2 \pi f) \big[ (\roll^4/\tildebr^4)^{2/k} +1 \big] + \xi$, where $\xi \in [0,1)$ such that $\bar{\ell} \in \mathbb{N}$.

\begin{figure}[t]
	\centering
	\raisebox{0\height}{\includegraphics[width=0.6\textwidth]{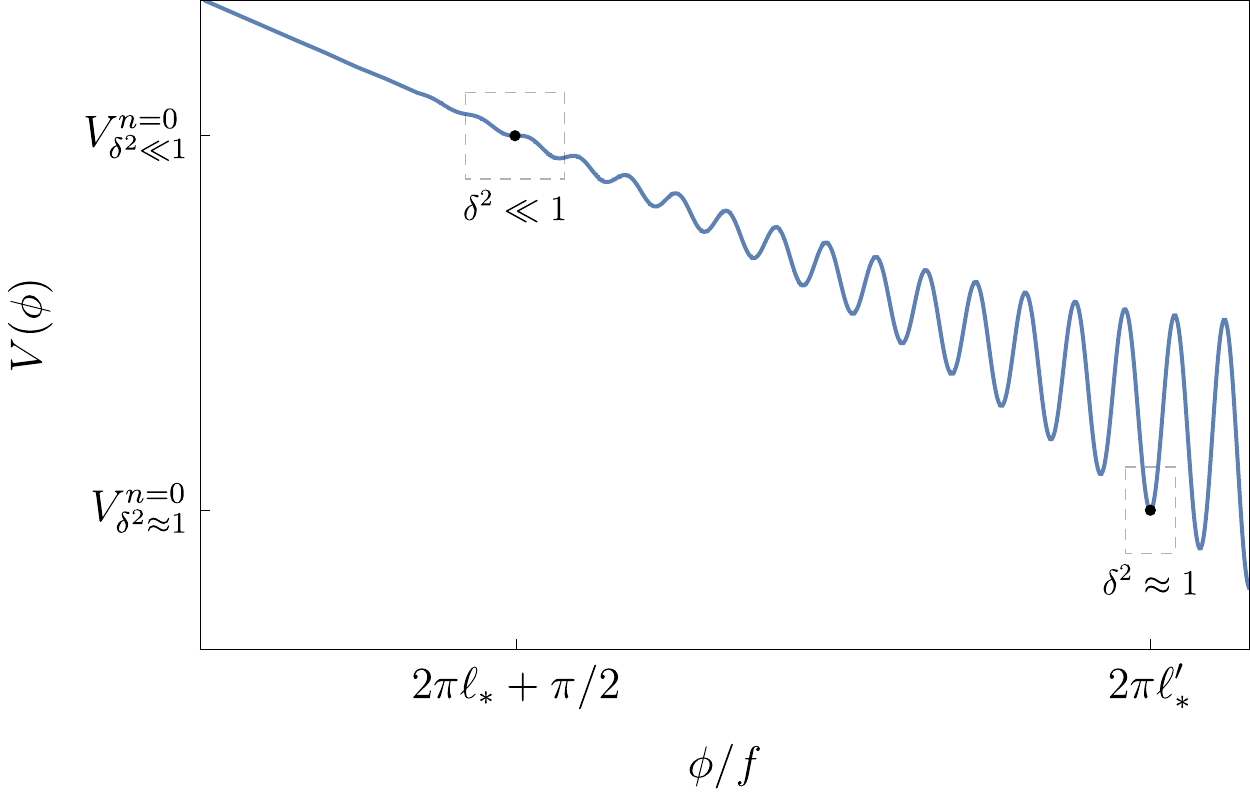}}
	\raisebox{-1\height}{\includegraphics[width=0.48\textwidth]{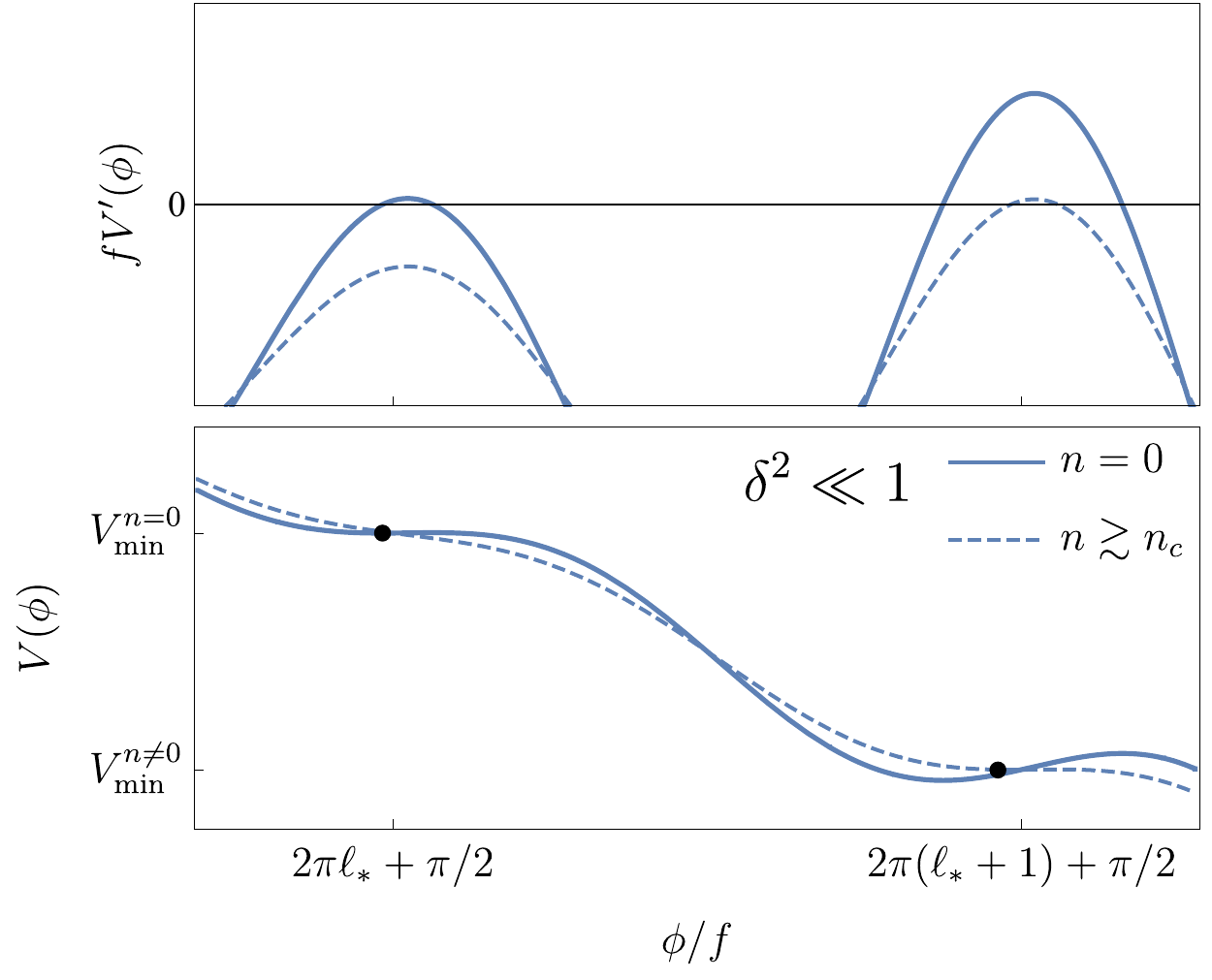}}
	\hspace{0.2cm}
	\raisebox{-1\height}{\includegraphics[width=0.48\textwidth]{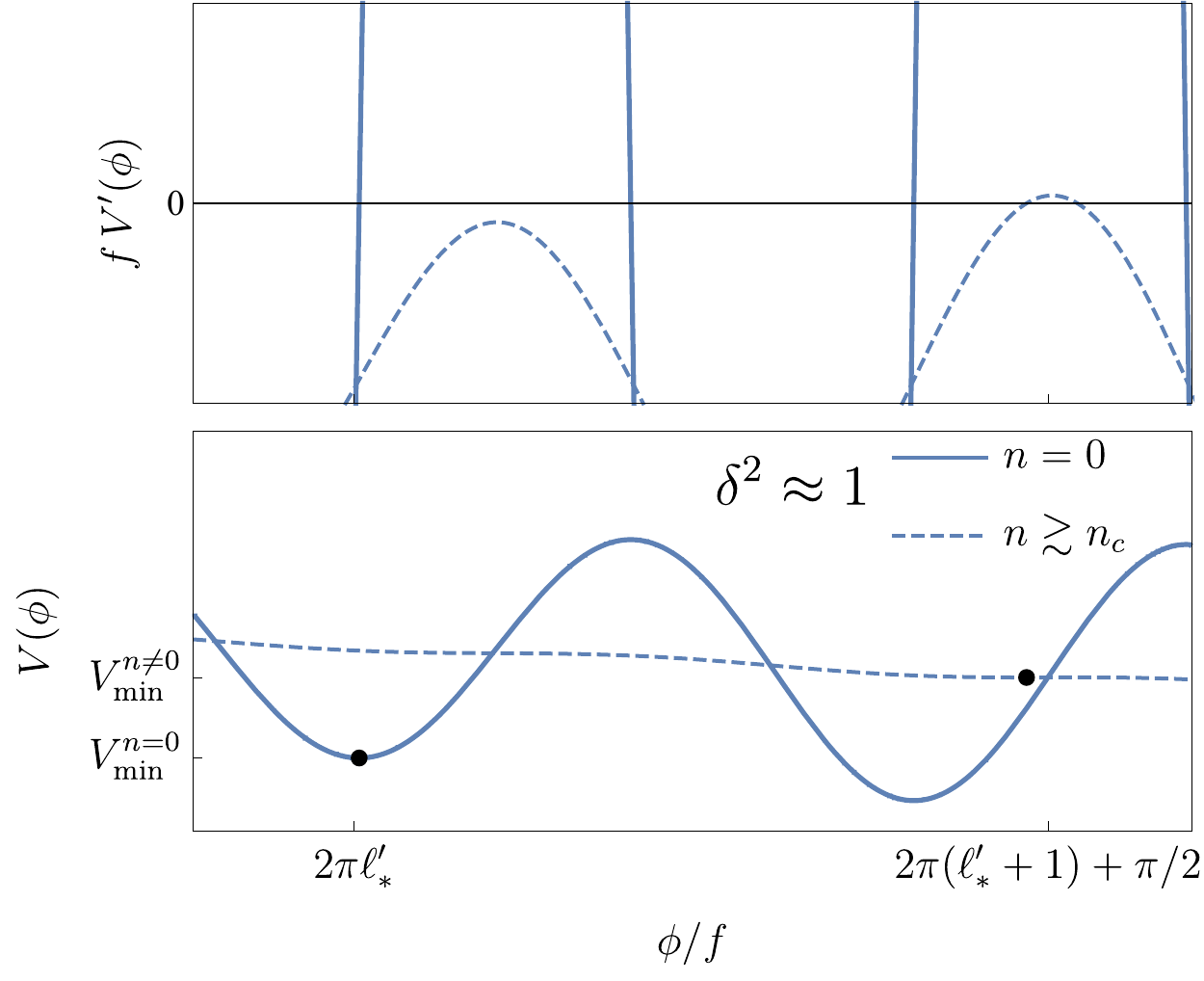}}
	\caption{Global view of the relaxion potential (upper panel) and zoomed in regions with shallow (lower left) and deep (lower right) minima. Also plotted in the lower panels the in-density potential for $n$ slightly larger than critical (dashed). The upper sub-panels show the derivative of the potential, both for $n=0$ and $n \gtrsim n_c$. The first minima at finite density are always shallow. }
	\label{fig:potentials}
\end{figure}

The relaxion landscape in \Eq{eq:Vgeneric} has two qualitatively different types of minima, depending on the relative size of the rolling and back-reaction terms, see \Fig{fig:potentials}. These can be parametrized by the variable~$\delta$
\beq
f V'(\phi_{\ellm}(\pi/2)) \simeq -\roll^4 +\br^4 \equiv \delta^2_{\ellm} \br^4 \,,
\label{eq:delta}
\eeq
where $V'$ is the derivative of the potential, here evaluated at the period $\ellm$. 
We note that $\delta_{\ellm}$ depends on the period, although to ease the notation we shall henceforth omit the subscript whenever unnecessary.
For the first periods of the potential in which a minimum is present, the parameter $\delta$ is small. This implies that these minima are \emph{shallow} \cite{Banerjee:2020kww}. Indeed, for $\delta^2 \ll 1$ the minimization condition $0 = f V'(\phi_{\ellm}(\thetam)) \simeq -\roll^4 +\br^4 \sin(\thetam)$ is satisfied at $\thetam \simeq \pi/2 - \sqrt{2} \delta$, which is very close to the local maximum (found at $\theta \simeq \pi/2 + \sqrt{2} \delta$), see the lower-left panel of \Fig{fig:potentials}. The mass of the relaxion in these minima is given by
\beq
\qquad m_{\phi}^2 \simeq \frac{\br^4}{f^2} \sqrt{2} \delta\,, \qquad ({\rm shallow})
\label{eq:massdelta}
\eeq
parametrically suppressed with respect to the usual expectation $m_{\phi}^2 \simeq \br^4/f^2$. In subsequent minima one finds $m_{\phi_{\ellm}}^2 = \sqrt{\ell_*} \, m_{\phi_{1}}^2$, where the value of $\delta^2$ corresponding to $\ell_* = 1$ is 
\beq
\delta_{\ellm = 1}^2 \simeq \frac{k \pi f}{\phi_c} \bigg( \frac{\tildebr^4}{\roll^4} \bigg)^{2/k} \left( \tfrac{1}{4}+\xi \right) \, .
\label{eq:delta1}
\eeq
The other type of minima we are interested in corresponds to $\delta^2 \approx 1$. These are \emph{deep} minima, since $\br \gg \roll$, see the lower-right panel of \Fig{fig:potentials}. The minimization condition yields $\thetam \simeq 1-\delta^2 \ll 1$, and the relaxion mass is simply
\beq
\qquad  m_{\phi}^2 \simeq \frac{\br^4}{f^2}\,. \qquad  ({\rm deep})
\label{eq:masstheta}
\eeq
Another quantity of phenomenological interest, which is markedly different between shallow and deep minima, is the height of the potential barrier,
\begin{align}
\barr &\simeq \begin{cases}
4 \sqrt{2} \br^4 \delta^3 \,, \qquad & ({\rm shallow}) \\
2 \br^4 \, . \qquad \qquad  & ({\rm deep})
\end{cases}
\label{eq:barrier}
\end{align}
The suppression of the barrier in the case of minima with $\delta^2 \ll 1$ implies that even a small perturbation of the potential can easily destabilize the relaxion, displacing it towards lower energy minima.

%%%%%%%%%%%%%%%%%%%%%%%%%%%%%%%%%%%%%%%%%%
%%%%%%%%%%%%%%%%%%%%%%%%%%%%%%%%%%%%%%%%%%
\section{The relaxion at finite density} \label{sec:denserelaxion}

Similar to temperature, finite density effects can have a strong impact on potentials of the sort of \Eq{eq:Vgeneric}. While a priori these corrections could modify the landscape in several ways, here we mainly focus on a decrease of the back-reaction term, since for the relaxion this constitutes the leading deformation in most circumstances.%
\footnote{In \Sec{sec:relaxiontc} we discuss a relaxion model in which the rolling term changes due to an electromagnetic background, as well as the modifications that this possibility introduces w.r.t.~what we present in this section.}
Indeed, in QCD-relaxion models, the size of the potential barriers is controlled by $\qcd$, which is known to linearly decreases with (small) baryon density \cite{Cohen:1991nk,Alford:1997zt}, a fact that has already been shown to affect the QCD axion \cite{Hook:2017psm,Balkin:2020dsr}. Alternatively, for non-QCD-relaxion models, we show in \Sec{sec:relaxionnoqcd} that it is the change of the Higgs VEV with density that leads to a reduction of the potential barriers. In addition, in \Sec{sec:dark} we speculate about the possibility of dense systems made of the dark baryons, where the back-reaction term would decrease in a similar fashion as in QCD.

The scenario where the size of the potential barriers depends on density, $n$, has been recently investigated in detail in \cite{Balkin:2021zfd}. Most of the discussion here parallels the one presented there. We summarize the main results and adapt the notation when necessary to match the relaxion potential.

We introduce the dimensionless quantity $\zeta$ to parameterize the relative change in the back-reaction term at finite density
\beq
\zeta(n)\equiv 1-\frac{\tildebr^4(n)}{\tildebr^4}   \,,
\label{eq:brn}
\eeq
with $\zeta(n) \geqslant 0$ and $\zeta(0)=0$. 
Let us now consider that in vacuum the relaxion sits at the minimum corresponding to some fixed period $\ellm$. 
One can then define a critical density, $n_c$, above which the effective back-reaction $\br$ at this minimum is no longer larger than the rolling term. This is implicitly given by
\beq
\zeta_c  \equiv \zeta(n_c) = \delta_{\ellm}^2 \, ,
\label{eq:zetac}
\eeq
where we recall that $\delta_{\ellm}^2 = 1 - \roll^4/\br^4$.
When the critical density is reached, the local minimum associated with $\br$ ceases to exist. In other words, the relaxion minimum corresponding to the period $\ellm$ (and obviously all the previous minima) is destabilized when $\zeta(n) > \delta_{\ellm}^2$.
Shallow minima, where $\delta^2 \ll 1$, are easily destabilized by density corrections, since $\zeta_c \ll 1$, while deep minima require $\zeta_c \approx 1$ in order to disappear, see \Fig{fig:potentials}. In the following we restrict our discussion to $\zeta(n) \leq 1$, leaving the discussion of the case where the barriers change sign to \App{sec:opposite}.

Hence at densities $n > n_c$ the minimum in which the relaxion resides in vacuum disappears. The number of periods between such a minimum and the first minimum of the in-medium potential is given by
\beq
N \equiv \ell_{* \, n} - \ell_{*} = \frac{\phi_c}{2 \pi f} \left[ \left( \frac{1-\zeta_c}{1-\zeta} \right)^{2/k} - 1 \right] \bigg( \frac{\roll^4/\tildebr^4}{1-\zeta_c} \bigg)^{2/k} + \xi \,,
\label{eq:N}
\eeq
where $\xi \in [0,1)$ such that $N \in \mathbb{N}$. Independently of $k$ ($k \neq 0$), $N$ scales with the difference $\zeta - \zeta_c$. In addition, $N$ scales with $\phi_c/f \gg 1$, thus as soon as $\zeta$ is above the critical value, the first in-density minimum is generically many periods beyond the one in vacuum.
Additionally, the first minimum at finite density is always shallow: it lies at $\theta_{* \, n} \simeq \pi/2$, and the mass of the scalar as well as the potential barrier are suppressed beyond the naive expectations, much like for the shallow minima in vacuum.

The change in the ground state energy between consecutive minima, be these shallow or deep, at zero or non-zero density, is always
\beq
\Delta \Lambda \simeq - 2 \pi \roll^4 \,.
\label{eq:lambda}
\eeq

%%%%%%%%%%%%%%%%%%%%%%%%%%%%%%%%%%%%%%%%%%
\subsection{Formation and escape of a bubble} \label{sec:bubble}

The disappearance of the in-vacuo minimum at supercritical densities leads to a non-trivial scalar profile, $\phi(r)$, developing within finite size systems such as stars. The characteristic scale of such a relaxion bubble is set by $\mu = \roll^2/f$. This implies that for stars whose core, defined as where $n(r) > n_c$, is larger than $\mu^{-1}$, the field is displaced from its value in vacuum by an amount $\Delta \phi(0) \gtrsim 2 \pi f$. Note that this means that inside the bubble, the relaxion has surpassed the field distance corresponding to one period of the potential in vacuum.
This gives rise to the possibility that a relaxion bubble, originally confined to dense system, expands beyond the core of the star, given that such a value of relaxion corresponds to a lower energy configuration also in vacuum.
In fact, if the gain in energy density is large enough to compensate for the increase of the surface tension of the bubble between the core and outside of the star, the relaxion bubble expands indefinitely.

The conditions for the formation and escape of a scalar bubble interpolating between two adjacent minima of the potential have been recently derived in \cite{Balkin:2021zfd}, for a quartic scalar potential with a tilt. In the neighbourhood of a given minimum of the relaxion potential, the same analysis applies. Intuitively, this is to be expected because the conditions for formation and escape simply follow from energy considerations. First, a scalar bubble associated with a field displacement of one period, that is $2 \pi f$, forms when the associated gain in energy density, $\epsilon = -\vev{\Delta \Lambda} \sim 2 \pi \roll^4$ (see \Eq{eq:lambda}) is enough to compensate for the field gradient $\tfrac{1}{2}\vev{\phi'^2} \sim (2 \pi f/\Rt)^2$, where $\Rt$ is the radius of the star's core. 
Second, in a finite-size dense system there is an additional contracting force acting on a bubble wall (of radius $R$), besides the usual expanding force associated with $\epsilon$ and the contracting force associated with its surface tension $2 \sigma/R$. This additional force is due to the increase of $\sigma$ with $R$ from the core to the outer edge of the star, $\sigma' = d \sigma/ d R \sim \Delta \sigma/\DRt$, where $\DRt$ is the size of the transition region from the core to the end of the star. Since for a sufficiently large bubble the surface-tension force becomes small, a scalar bubble can expand beyond the confines of the star if $\epsilon \gtrsim \sigma'$.

For a shallow relaxion minimum ($\delta^2 \ll 1$), the change in the wall's tension is negligible, since already in vacuum the potential barrier separating the two adjacent minima is very small, \Eq{eq:barrier}. Therefore, when a shallow relaxion bubble forms, it generically escapes from the star as well. The condition for this to happen is
\beq
\Rt \gtrsim \frac{f}{\roll^2} \,. \qquad ({\rm shallow})
\label{eq:condshallow}
\eeq
Instead, for a deep relaxion minimum ($\delta^2 \approx 1$), the change in the wall's tension is significant, going from being dominated by the gradient energy at the core, to being dominated by the large potential barrier in vacuum. This implies that the escape condition is stronger than the condition for formation. The former reads
\beq
\DRt \gtrsim \frac{f}{\roll^2} \frac{1}{\sqrt{1-\delta^2}} = \frac{f \br^2}{\roll^4} \, .\qquad ({\rm deep})
\label{eq:conddeep}
\eeq
As expected, it is generically much more difficult for a bubble connecting deep minima to escape from the star than for a shallow bubble. Furthermore, recall that in order to destabilize a deep minimum much larger densities are needed than in the shallow case.

In practice, we take both the size of the core $\Rt$, and the size of the transition region from the core to the end of the star $\DRt$, to be of the same order as the whole size of the star, $\Rs \sim \Rt \sim \DRt$. This is justified for density profiles where the core density is not accidentally close to the critical one.
In this regard, notice that different types of stars span not only a range of radii but a range of core densities as well, so cases in which $\Rt, \, \DRt \ll \Rs$ are not generic. 

In the following we consider main-sequence stars like the Sun, white dwarfs (WD), or neutron stars (NS).
We take the following as typical radii%
\footnote{The simple estimates for $\Rns$ and $\Rwd$, are obtained by equating (Fermi-degeneracy) kinetic and gravitational energy densities, e.g.~for a neutron star $\rho_k \sim m_n n_n \sim m_p^4$ and $\rho_g \sim m(r)m_n n_n/8\pi\mplanck^2r \sim m_p^8 r^2/8\pi\mplanck^2$, where $n_n \sim m_n^3$ is the (constant) neutron density and $m(r) \sim m_n n_n r^3$ is the enclosed mass.\label{foot:estimates}} 
%Note estimate is not good for neutron star density: QCD interactions and gravity are important.
\begin{align}
\Rns & \equiv \sqrt{8\pi} \mplanck/m_p^2 \approx 2.7 \km \, , \nonumber \\
\label{eq:refrs}
\Rwd & \equiv \Rns (m_p/m_e) \approx 5 \times 10^3 \km \, , \\
R_\odot & \approx 7 \times 10^5 \km \, . \nonumber
\end{align}
For the baryonic densities of these types of stars, which are the most relevant for relaxions, we take as typical (average) values
\begin{align}
\nns & \equiv n_0 \approx 0.16 / {\rm fm}^3 \approx 1.3 \times 10^6 \MeV^3 \, , \nonumber \\
\label{eq:refns}
\nwd & \equiv  2.8 \times 10^7 \, {\rm g}/{\rm cm}^3 \times 1/m_p \approx 0.13 \MeV^3 \, , \\
n_\odot & \approx 1.5 \, {\rm g}/{\rm cm}^3 \times 1/m_p \approx 7 \times 10^{-9} \MeV^3 \, , \nonumber
\end{align}
where $n_0$ is the nuclear saturation density. Note that for white dwarfs we have in fact taken $\nwd = m_e^3$, as determined by Fermi degeneracy, since this matches their mean density well \cite{Shapiro:1983du,Glendenning:1997wn}.

Before moving to the phenomenological consequences of escaping bubbles for specific relaxion models, several additional comments are in order:

The conditions in \Eqs{eq:condshallow}{eq:conddeep} hold under the assumption that the typical reaction time of the relaxion field (set by $\mu^{-1} = f/\roll^2$) is much faster than the time it takes for a forming star to develop a core (in which $n > n_c$) of size $\Rt \gtrsim \mu^{-1}$. For stellar processes where this is not the case, the formation of a scalar bubble takes place suddenly instead of in the nearly static fashion we have assumed. As discussed in \cite{Balkin:2021zfd}, this can modify the escape condition of the scalar bubble, although only in the case that $\Rt \gg \mu^{-1}$.

Irrespective of the time evolution, if the stars grows very large compared to $\mu^{-1}$, very large field displacements (w.r.t.~to where the relaxion resides in vacuum) are energetically allowed inside the bubble. Indeed, at a fixed core density such that $\zeta$ is not accidentally close to criticality, a very large core $\Rt \sim \sqrt{N}\mu^{-1} \gg \mu^{-1}$ allows for the relaxion to move by many periods $N \gg 1$, see \Eq{eq:N}.
As explained in \App{sec:Nlarge}, such a large relaxion bubble has the effect of helping the standard bubble connecting two adjacent minima (i.e.~for which $\Delta \phi(0) \sim 2 \pi f$) to escape from the star. In fact, such a bubble escapes independently of the density profile, regardless of how fast density decreases towards the outer edge of the star. 
In this case the conditions \Eqs{eq:condshallow}{eq:conddeep} read the same, only with $\Rs$ instead of $\Rt$ and $\DRt$; they simply encode the requirement for a standard relaxion bubble to expand once it is outside of the star, see \Eqs{eq:Rsminsub}{eq:condsub}.
If one bubble is able to escape, the new relaxion minimum in vacuum becomes the one associated with the next period, i.e.~$\ellm + 1$.
Interestingly, since $N = \ell_{* \, n} - \ell_{*} \gg 1$, such a new minimum is also unstable inside the dense system. This then implies that another bubble, within which this time the relaxion sits at the minimum $\ellm + 2$, will generically be able to escape as well, and so on until the escape condition is no longer satisfied.

Finally, let us note that our whole discussion relies on the assumption that the density profile is treated as a background field that does not receive a large back-reaction from the formation and expansion of a relaxion bubble. In \App{app:environment} we discuss the interactions of the relaxion (non-trivial configurations) with the density profile, thereby justifying this treatment.

%%%%%%%%%%%%%%%%%%%%%%%%%%%%%%%%%%%%%%%%%%
%%%%%%%%%%%%%%%%%%%%%%%%%%%%%%%%%%%%%%%%%%
\section{Bounds on relaxions}\label{sec:relaxionbounds}

The relaxion bubbles discussed in the previous section are born along with the stars that seed them. Therefore, if the conditions for the bubble to expand beyond the dense system are met, a phase transition in the universe to a new relaxion vacuum can take place whenever the right type of stars are formed. The first stars were born at redshifts $z = O(10)$, along with galaxy formation.
Under the assumption that the phase transition is completed before redshifts $z \sim 1$, an immediate consequence is that the vacuum energy of the universe, $\Lambda$, is not the same at recombination than in the late universe. 
A rough experimental bound on such a change was derived in \cite{Balkin:2021zfd},
\beq
- \Delta \Lambda/\Lambda_0 \lesssim 10^2 \, ,
\label{eq:lambdabound}
\eeq
with $\Lambda_0 \approx (2.3 \, \rm{meV})^4$ the value of the vacuum energy inferred from the standard cosmological model.
Since a relaxion bubble is associated with a gain in ground state energy given by $- \Delta \Lambda \sim 2 \pi \roll^4$, this leads to constraints on relaxion models.
Indeed, recalling that for a shallow minimum the condition for formation and escape \Eq{eq:condshallow} can be rewritten as $\roll^2 \gtrsim f/\Rs$, the occurrence of a phase transition implies
\beq
- \Delta \Lambda \gtrsim 2 \pi \left( \frac{f}{\Rs} \right)^2 \approx 2 \times 10^6 \, \Lambda_0 \left( \frac{f}{10 \TeV} \right)^2 \left( \frac{R_\odot}{\Rs} \right)^2 \,,
\label{eq:lambdachange}
\eeq
where we have taken the Sun's radius as a reference value for the largest stars considered to derive our constraints, and $f = 10 \TeV$ as a large enough decay constant such that the cutoff associated with the relaxion potential, $M \lesssim 4 \pi f$, is sufficiently above the electroweak scale. 
Note that for a deep bubble to escape, an even larger energy difference is needed, by a factor $(1-\delta^2)^{-1} \gg 1$, see \Eq{eq:conddeep}.
The possibility of such a phase transition is clearly ruled out by cosmological data.
Still, it is interesting to note that if we were to consider bubbles nucleated by the largest stars observed so far, with radii $\Rs \sim 10^3 R_\odot$, or by large non-standard astrophysical objects, associated for instance with some beyond the SM relic species, e.g.~dark matter (see \Sec{sec:dark}), then a phenomenologically viable late-time phase transition could have taken place. In this regard, it would be interesting to perform a detailed assessment of the associated cosmological and astrophysical signatures.

Besides, in relaxion models the change of minimum also implies a change in the Higgs VEV, for which there exist cosmological (and astrophysical) constraints as well. However, let us note right away that the relative change of the electroweak scale between minima is much smaller than the change in the vacuum energy: $(\Delta h^2/v^2)/(|\Delta \Lambda|/\Lambda_0) = \Lambda_0/ \lambda c v^2 M^2 \ll 1$, where we have used \Eq{eq:higgschange}. 
Nevertheless, since this is one of the trademarks of relaxion models compared to other landscapes, let us briefly review the bounds.
There are significant constraints on a different value of the Higgs VEV during BBN, $|\Delta h/v| \lesssim 10^{-2}$ where $\Delta h = h - v$ \cite{Yoo:2002vw}.
In addition, it has been recently argued that SN explosions can only happen if $h$ is below a factor of a few away from $v$ \cite{DAmico:2019hih}. 
While these constraints (the one from BBN in particular) could be violated if the universe underwent a change of relaxion minimum at star formation, as shown above the associated change in vacuum energy always yields a more or equally stringent constraint.

In the remainder of this section, we work out the specifics of how a non-vanishing SM matter density (or an electromagnetic background) affects the potential of some benchmark relaxion models, and re-express the conditions for the formation and escape of bubbles in terms of their parameters.

%%%%%%%%%%%%%%%%%%%%%%%%%%%%%%%%%%%%%%%%%%
\subsection{QCD relaxion} \label{sec:relaxionqcd}

The most economic origin of the relaxion periodic term is low-energy QCD dynamics, in which case we identify the relaxion with the QCD axion. The dependence on the Higgs VEV arises from the well-known dependence of the axion potential on the quark masses. This leads us to identify $\tildebr$ in \Eq{eq:Vgeneric} as well as $\br$, the effective size of the periodic term at the minimum where the relaxion sits in vacuum, as
\beq
\tildebr^4 = \qcd^4 \frac{M}{v \sqrt{\lambda}} \, , \quad \br^4  = \qcd^4 \frac{h}{v} \, .
\label{eq:brqcd}
\eeq
Note that if a seeded phase transition took place, the relaxion would not sit at the same minimum today than right before star formation. Nevertheless, in the following we conservatively fix $h = v \approx 246 \GeV$, since any minimum prior to star formation with a smaller $h$ would necessarily be shallower than the present one, making it easier for the transition to take place.
The exponent of the function $F(\phi)$ in \Eq{eq:F} is determined as well, $k = 1$, since the dependence of the back-reaction term on the Higgs VEV is linear. Note that the QCD(-axion) scale is $\qcd^4 \simeq m_\pi^2 f_\pi^2/4$.
Finally, the value of $\delta$, which determines if the relaxion minimum is shallow or deep, depends on the relative size of the rolling term, $\roll^4 = g M^3 f$, according to \Eq{eq:delta},
\beq
\delta^2 = 1 - g \frac{M^3 f}{\qcd^4} \, .
\label{eq:deltaqcd}
\eeq
In this regard, let us note that, as advanced, for the first minima of the potential $\delta$ is always a small parameter as long as the scanning of the Higgs VEV is sufficiently precise. At the first minimum,
\beq
\delta_{\ellm = 1}^2 \simeq \frac{\pi \qcd^4}{\lambda v^2 M^2} \left( \tfrac{1}{4}+\xi \right) \ll 1 \, , \quad \xi \in [0,1) \, ,
\label{eq:delta1qcd}
\eeq
while for all the subsequent minima $\delta_{\ellm}^2 = \ellm \delta_1^2$. Since it has no actual significance, from now on we set $\xi = 0$.

Once all the relevant parameters of our potential have been identified, let us consider the fate of the relaxion bubbles, starting with shallow minima, $\delta^2 \ll 1$.
This case should be mostly considered as illustrative, since the value of the relaxion at the minimum is displaced from a multiple of $2 \pi f$ by approximately $\pi/2$, thus the strong CP angle is also $\thetacp \simeq \pi/2$, which is experimentally ruled out. 
Keeping this in mind, we can compute the dependence of the back-reaction term, or equivalently $\zeta$ in \Eq{eq:brn}, on the baryon density $n_b$ by means of the Hellmann-Feynman theorem, as explained in e.g.~\cite{Balkin:2020dsr},
\beq
\zeta(n_b) \simeq \frac{\sigma_{\pi N} n_b}{m_\pi^2 f_\pi^2} \, .
\label{eq:zetaHF}
\eeq
This holds for densities below a few times nuclear saturation, and where $\sigma_{\pi N} \approx 45 \MeV$ is known as the pion-nucleon sigma term.
In turn, since the critical value of $\zeta$ for which the relaxion can classically move is given by $\zeta_c = \delta^2$, we find that a proto-bubble can start forming if
\beq
n_b > \frac{\ellm}{M^2} \frac{\pi \qcd^8}{\sigma_{\pi N} \lambda v^2} \approx 1 \times 10^{-8} \MeV^3 \left( \frac{1 \TeV}{M/\sqrt{\ellm}} \right)^2 \, ,
\label{eq:ncqcd}
\eeq
that is if densities are larger than $3 \, {\rm g}/{\rm cm}^3 \times 1/m_p$.
This is a very low critical density, found not only in neutron stars and white dwarfs, but in the Sun as well. 
The densities reached in these systems then set the minimum value of $M/\sqrt{\ellm}$ that is excluded if the bubble eventually fully forms and escapes the star.
The corresponding condition is given in \Eq{eq:condshallow}, which for the QCD-relaxion reads
\beq
\Rs \gtrsim \frac{f}{\qcd^2} \, ,
\label{eq:rscqcd}
\eeq
where we have taken $\Rt \sim \Rs$ as argued in \Sec{sec:bubble}, and traded $\roll$ with $\qcd$ given that $\delta^2 \ll 1$. Using the reference radii and densities quoted in \Eqs{eq:refrs}{eq:refns}, we arrive at the following excluded values for the relaxion decay constant and cutoff
\begin{align}
{\rm NS:} & \quad f \lesssim 3 \times 10^{-2} \mplanck \, , \quad M/\sqrt{\ellm} \gtrsim 1 \times 10^{-4} \GeV \, , \nonumber \\
\label{eq:qcdshallow}
{\rm WD:} & \quad f \lesssim 63 \, \mplanck \, , \;\; \quad \qquad M/\sqrt{\ellm} \gtrsim 0.3 \GeV \, , \qquad \qquad (\textrm{QCD; shallow})\\
\odot: & \quad f \lesssim 9 \times 10^3 \mplanck \, , \;\; \quad M/\sqrt{\ellm} \gtrsim 1.4 \TeV \, . \nonumber
\end{align}
Therefore, while recalling that the shallow QCD relaxion is already ruled out by a too large $\thetacp$, we find that classical rolling and escape would happen for nearly all values of $f$ and $M/\sqrt{\ellm}$. Note in fact that for both white dwarfs and main-sequence stars the upper bounds on $f$ are above $\mplanck$, and that for neutron stars and white dwarfs the lower bound on $M$ is not larger than the electroweak scale.

The situation is markedly different for deep minima, in particular since we must require $\thetacp \lesssim 10^{-10}$, which then fixes $1-\delta^2$ to be as small at the minimum in question. Since $\zeta_c = \delta^2$, this implies that the QCD barriers at finite density, $\qcd^4 (1-\zeta(n))$, would need to nearly disappear for the relaxion to be able to classical move to the next minimum. Such large densities, if attainable at all inside neutron stars, are certainly beyond perturbative control and thus the linear approximation used to derive \Eq{eq:zetaHF} is not applicable.
However, since in the cores of neutron stars densities could be higher than ten times nuclear saturation density \cite{Lattimer:2004pg}, it has been long been hypothesised that new phases of QCD, such as kaon condensation or color-superconductivity, could take place there, see e.g.~\cite{Balkin:2020dsr} and references therein. As shown in that work, this raises the possibility that, while remaining finite, the periodic potential flips sign. As explained in \App{sec:opposite} (see also \cite{Hook:2019pbh}), this would lead to relaxion condensation with $\Delta \phi(0) = \pi$, assuming a small rolling region. Such a type of bubble would remain confined inside the dense system.

Still, an interesting option remains that such a change of phase of strongly interacting matter, being controlled by QCD dynamics, happens very fast compared to the reaction time of the relaxion. In the case of a deep minimum this reaction time is prologned compared to a shallow one, $\mu^{-1} = f/\roll^2 = \thetacp^{-1/2}f/\qcd^2$.
Then, as discussed in \cite{Balkin:2021zfd}, the kinetic energy that the field acquires after the sudden change of its potential could be enough to overcome the (flipped) barriers and to create a relaxion bubble with $\Delta \phi(0) \gg 2 \pi f$. This facilitates the escape of a standard $2 \pi f$ bubble, as discussed above (see also \App{sec:Nlarge}).
With our current knowledge of QCD at such extreme densities we cannot assert whether this is the right picture. Nevertheless, if it were, a phase transition would take place if the condition \Eq{eq:condsub} is satisfied
\beq
f < \sqrt{\thetacp} \qcd^2 \Rns \approx 8 \times 10^{11} \GeV \left(\frac{\thetacp}{10^{-10}} \right)^{1/2} \, . \qquad (\textrm{QCD; deep})
\label{eq:qcddeep}
\eeq

%%%%%%%%%%%%%%%%%%%%%%%%%%%%%%%%%%%%%%%%%%%
\subsection{Non-QCD relaxion} \label{sec:relaxionnoqcd}

The correlation between the relaxion selection of the electroweak scale and of $\thetacp$, i.e.~between the electroweak hierarchy and the strong CP-problem, can be broken by positing that dynamics other than QCD is responsible for the generation of the periodic back-reaction term \cite{Graham:2015cka}. Such a non-QCD strong sector must still couple to the relaxion in such a way as for the amplitude of the barriers to depend on the Higgs VEV. 
Experimental constraints on new electroweak-charged degrees of freedom that get mass from electroweak symmetry breaking motivates that such a dependence is quadratic, instead of the linear dependence of the QCD scale (see however \Sec{sec:relaxiontc}). Therefore, we identify our potential parameters in \Eq{eq:Vgeneric} as
\beq
\tildebr^4 = \noqcd^4 \frac{M^2}{\lambda v^2} \, , \quad \br^4  = \noqcd^4 \frac{h^2}{v^2} \, , \quad k = 2 \, ,
\label{eq:brnoqcd}
\eeq
where $\noqcd$ is the new confinement scale, analogous to $\qcd$ in \Eq{eq:brqcd}. In order for the size of the barriers to be naturally dominated by the Higgs VEV squared, the condition $\noqcd^2 \lesssim 4 \pi v^2$ must be required \cite{Graham:2015cka,Espinosa:2015eda,Flacke:2016szy,Fonseca:2019lmc}.
In parallel with the QCD-relaxion, the value of $\delta$ at a given minimum is determined by the relative size of the rolling term and $\noqcd^4$, i.e.~\Eq{eq:deltaqcd} with $\qcd \to \noqcd$. The first minima of the landscape are always shallow, since $\delta_1^2 \simeq \pi \noqcd^4/2\lambda v^2 M^2 \ll1$ for $\noqcd^2 \lesssim 4 \pi v^2$ and $M \gg 4 \pi v$.\\

The dependence of the back-reaction term on the (SM) matter density in this case is indirect, stemming from a change in the Higgs VEV. This is due to the coupling of the Higgs field to fermions, $\mathcal{L} \supset - \tfrac{1}{\sqrt{2}} \, y_\psi h \bar{\psi} \psi $, which in a (non-relativistic) $\psi$ background, $\vev{\bar \psi \psi} \simeq \vev{\bar \psi \gamma_0 \psi} = n_\psi$, displaces its VEV from its value in vacuum.
The small relative displacement with respect to \Eq{eq:higgsvev} is given, at leading order in $n_\psi$, by
\beq
\delta h^2 (n_\psi) = \frac{y_\psi}{\sqrt{2}} \frac{n_\psi}{\lambda v^3} \, ,
\label{eq:dhiggsvevn}
\eeq
where we have evaluated $h = v$.
The change $h^2 \to h^2 ( 1 + \delta h^2)$ is then responsible for the density dependence of the relaxion potential. In this regard, note that both the rolling and back-reaction terms are affected, since both of them are in fact quadratic in the Higgs, see \Eq{eq:higgs} and \Eq{eq:brnoqcd} respectively.
Nevertheless, it is easy to see that the leading effect is associated with the latter since, while the Higgs contribution to the barriers is the leading piece, it is a subleading one for the linear slope as long as $v^2/M^2 \ll 1$.

The most relevant densities to consider, as in the case of the QCD-relaxion, are baryonic, since these are usually the largest (in particular in the cores of neutron stars) and the coupling of nucleons to the Higgs is significant $y_N = \sigma_{\pi N}/v$, where $N = n, p$. In neutron stars, besides neutrons and protons, leptons are present as well. Charge neutrality implies $n_p + n_e + n_\mu = 0$, where note that due to the highly energetic Fermi surface of the electron, $\beta$-equilibrium not only implies $\mu_n = \mu_p + \mu_e$ but $\mu_e = \mu_\mu$ as well, implying a non-vanishing muon density (for $\mu_\mu > m_\mu$). This is interesting since the coupling of muons to the Higgs, $y_\mu = m_\mu/v$, is twice as large as to nucleons. In the outer layers of neutron stars, in white dwarfs and main-sequence stars, baryon densities become once again the most important, given the small coupling of electrons to the Higgs.
We therefore focus on the effects of a non-vanishing $n_b$. Still working in the linear approximation, the decrease of the non-QCD barriers is encoded as
\beq
\zeta(n_b) \simeq \sqrt{2} \frac{\sigma_{\pi N} n_b}{m_h^2 v^2} \, ,
\label{eq:zetanoqcd}
\eeq
where we have written it in terms of the physical Higgs mass, $m_h^2 = 2 \lambda v^2$, to make apparent the similarity with \Eq{eq:zetaHF}.
A relaxion bubble can then classically form if densities satisfy the following condition
\beq
n_b > \frac{\ellm \noqcd^4}{M^2} \frac{\pi v^2}{\sqrt{2} \sigma_{\pi N}} \approx 3 \times 10^{-3} \MeV^3 \left( \frac{1 \TeV}{M/\sqrt{\ellm}} \right)^2 \left( \frac{\noqcd \simeq \roll}{1 \MeV} \right)^4 \, .
\label{eq:ncnoqcd}
\eeq
Even though finite density effects are relatively suppressed in the case of the non-QCD relaxion, the required critical densities are sufficiently small, for large cutoffs or small back-reactions, that they can be found from neutron stars to the Sun. Furthermore, relaxion bubbles will only form for shallow minima, $\zeta_c = \delta^2 \ll 1$, since values of $\zeta(n_b)$ close to unity, which are required to destabilize deep minima, can never be achieved (this would require exorbitant densities, of order $m_h^2 v^2/\sigma_{\pi N} \sim 10^{19} \MeV^3$).

\begin{figure}[t]
	\centering
	\includegraphics[width=1.0\textwidth]{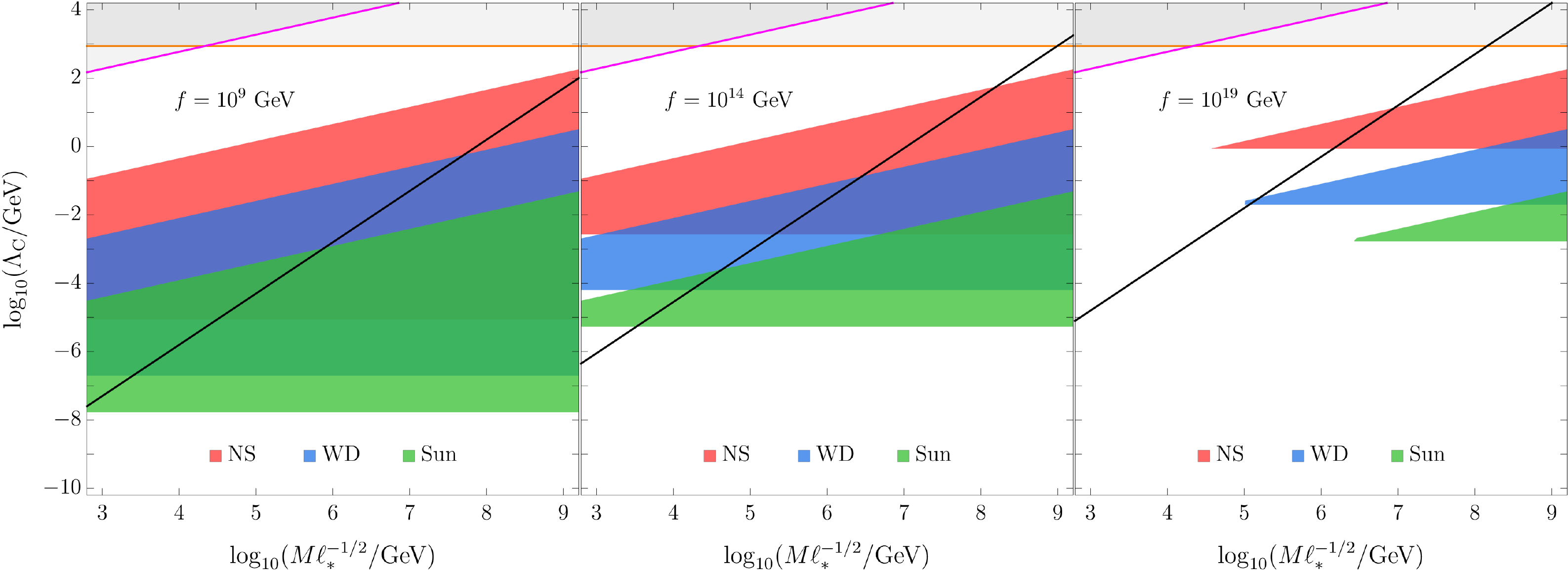}	
	\caption{$(M/\sqrt{\ellm},\noqcd)$-regions excluded by the formation and escape of a non-QCD relaxion bubble induced by neutron stars (red), white dwarfs (blue), and the Sun (green), in the case of shallow minima, $\delta^2 \ll 1$ or equivalently $\roll \simeq \noqcd$, and for $f = 10^{9}, \, 10^{14}, \, 10^{19} \GeV$ (left, middle, right panel, respectively). The grey, shaded region is excluded by the requirements $\noqcd^2 \lesssim 4 \pi v^2$ (orange line) and $\Delta h^2 < v^2$ (magenta line), while the region of parameter space preferred by relaxation during inflation lies above the diagonal black line. Note that the latter line depends on a model-dependent period of dynamical evolution, which our constraints are independent of. The three lines are drawn taking $\ellm = 1$.}
	\label{fig:boundsroll}
\end{figure}

The condition for a shallow bubble to fully form and escape the star simply reads
\beq
\Rs \gtrsim \frac{f}{\noqcd^2} \, , \qquad (\textrm{non-QCD; shallow})
\label{eq:rscnoqcd}
\eeq
where recall that for $\delta^2 \ll 1$ we have $\roll \simeq \noqcd$.
The conditions in \Eqs{eq:ncnoqcd}{eq:rscnoqcd}, which if satisfied imply a late-time phase transition at odds with experiment, give rise to non-trivial constraints on the parameter space of the non-QCD relaxion. These are qualitatively different and generically stronger than those dependent on the period of dynamical evolution; namely $\roll^4 > M^6 f/\sqrt{12} \pi \mplanck^3$ for relaxation during inflation, associated with the requirement of classical evolution of the field along with the energy density associated with the relaxion being a subdominant component \cite{Graham:2015cka,Espinosa:2015eda,Flacke:2016szy,Fonseca:2019lmc}.
A related but different discussion of chameleon effects relevant for dark matter direct detection experiments have been recently presented in \cite{Budnik:2020nwz}.
We show our constraints in \Figs{fig:boundsroll}{fig:boundsg}, in the planes $(M/\sqrt{\ellm},\noqcd)$, and $(M,g)$ for $\ellm = 1$, respectively. These are for three different values of the relaxion decay constant, $f = 10^{9}, \, 10^{14}, \, 10^{19} \GeV$. 
In both planes, it is evident that the lower boundary of the excluded (shaded) regions extends to smaller values of either $\noqcd \simeq \roll$ or $g$ as $f$ is taken smaller, since it is easier for the relaxion bubble to fit inside a given type of star, \Eq{eq:rscnoqcd}. In turn, as $f$ is taken larger, either $\noqcd$ or $g$ must be larger for the bubble to be able to form, which then requires higher densities, \Eq{eq:ncnoqcd}; this is why the constrains from less dense stars become comparatively weaker.
Note that although the plots are cut at $M = 10^9 \GeV$, the constraints actually extend up to $M \lesssim 4 \pi f$ in each case.
Let us also point out that if the theoretical expectation that $f < \mplanck$ is accepted, the constraints for $f = 10^{19} \GeV \approx \mplanck$ can be considered as absolute, meaning the corresponding parameter space is excluded for any (possible) value of the axion decay constant.
Finally, considering larger stars with enough density would enlarge the excluded regions. In the case of the (green) region associated with main-sequence stars, once the inequality \Eq{eq:lambdabound} is saturated, more refined experimental constraints on changes in the energy budget of the universe or from other observables, would be needed. The investigation of these detailed bounds is beyond the scope of this work.

\begin{figure}[t]
	\centering
	\includegraphics[width=1.0\textwidth]{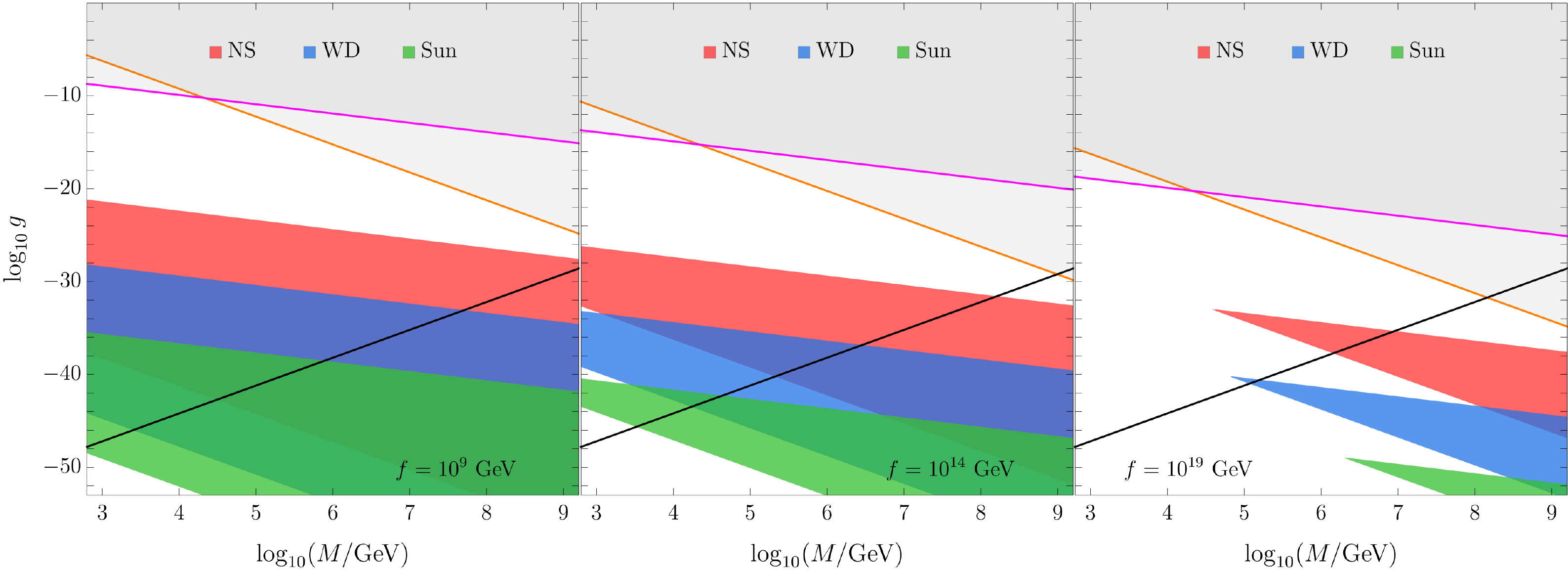}
	\caption{$(M,g)$-regions excluded by the formation and escape of a non-QCD relaxion bubble induced by neutron stars (red), white dwarfs (blue), and the Sun (green), in the case of shallow minima, $\delta^2 \ll 1$ or equivalently $\roll \simeq \noqcd$, and for $f = 10^{9}, \, 10^{14}, \, 10^{19} \GeV$ (left, middle, right pannel). The grey, shaded region is excluded by the requirements $\noqcd^2 \lesssim 4 \pi v^2$ (orange line) and $\Delta h^2 < v^2$ (magenta line), while the region of parameter space preferred by relaxation during inflation lies above the diagonal black line. Note that the latter line depends on a model-dependent period of dynamical evolution, which our constraints are independent of. Recall that $g = \roll^4/ M^3 f$ and we have taken $\ellm = 1$ in \Eq{eq:ncnoqcd}.}
	\label{fig:boundsg}
\end{figure}

%%%%%%%%%%%%%%%%%%%%%%%%%%%%%%%%%%%%%%%%%%%
\subsection{Technicolored relaxion} \label{sec:relaxiontc}

For the QCD and non-QCD relaxion models, the most important density deformation of their respective landscapes is in the form of a smaller back-reaction term. Now we wish to point out that in general this is not the only possibility. In this section we present a scenario in which the leading effect is due to a change in the rolling term. Furthermore, this change is induced not by background matter but by the electromagnetic fields surrounding a spinning neutron star.

Another variant of the relaxion model involves a technicolor-like sector which provides an additional source of electroweak symmetry breaking on top of the elementary Higgs.
While this sector, irrespective of the relaxion, is severely constrained experimentally (by electroweak precision data, Higgs coupling measurements and resonance searches at the LHC), it is not yet ruled out \cite{Chang:2014ida,Harnik:2016koz}.
Exactly like for the QCD axion, the coupling of the relaxion to the topological charge of this new confining sector gives rise to the periodic potential term \cite{Flacke:2016szy}, with the analog of $\qcd$ given by
\beq
\tc^4 \simeq 4 \pi v'^3 m_U \, ,
\label{eq:tcscale}
\eeq
where $v'$ is the electroweak-breaking order parameter of the technicolor (TC) sector. The electroweak scale is given by $v^2 = v'^2 + h^2$ and $m_U = y_U h/\sqrt{2}$ is the lightest techniquark mass, linearly proportional to the elementary Higgs VEV. The parameters of the relaxion potential are then identified as $\roll^4 = g M^3 f$ (like in all the previous models), while $\tildebr$ and $\br$ are similar to the QCD relaxion, \Eq{eq:brqcd}, with the following replacements
\beq
\qcd \to \tc\,, \;\;\; v \to \sqrt{v^2 - v'^2}\,.
\eeq
Due to the aforementioned experimental constraints, $v'$ cannot be large, $v' \lesssim 70 \GeV$, nor very small either, since the masses of the TC resonances are expected below $4 \pi v'$  \cite{Chang:2014ida}.

It is crucial for our analysis that the relaxion in this model has a large coupling to photons. 
Just like for the QCD axion, this coupling is a consequence of the (model-independent) coupling to the technigluons as well as the (model-dependent) electromagnetic anomaly,
\beq
\frac{g_{\phi \gamma \gamma}}{4}  \frac{\phi}{f} F_{\mu \nu} \tilde{F}^{\mu \nu} \, , \quad g_{\phi \gamma \gamma} = c \left(\frac{\alpha}{2 \pi}\right) \, ,
\label{eq:tcem}
\eeq
where $c$ is a model-dependent constant.
Such an interaction, which is not suppressed by the small shift-symmetry breaking parameter $g$, has significant implications for the fate of this relaxion model, in particular because of the existence of strong electromagnetic fields surrounding rapidly rotating neutron stars (magnetars/pulsars).%
\footnote{By considering a rotating star we are departing from our main assumptions concerning the characteristics of the system, as described in \cite{Balkin:2021zfd}, in particular spherical symmetry and (near) time-independence. However, we expect such departures to neither have a large impact on our qualitative description, nor to change the order of magnitude results we derive.}
Indeed, in such an environment the linear term in the relaxion potential receives an additional contribution, $\roll^4 \to \roll^4 (1+ \eta)$, where
\beq
\eta = \frac{g_{\phi \gamma \gamma} \vec{E} \cdot \vec{B}}{\roll^4} \, .
\label{eq:deltaroll}
\eeq
The electric and magnetic fields depend on the intrinsic properties of the star as well as on space-time, in a similar fashion as the (baryonic) density profiles that were considered in our previous examples.
However, in contrast to the case of a dense system of finite size, here the electromagnetic background extends to infinity (i.e.~much beyond the surface of the star). 
This implies, for instance, that the radius at which classical rolling is allowed is potentially much larger than $\Rs$.
As explained in \Sec{sec:bubble} (see \cite{Balkin:2021zfd} for a more detailed discussion) it is the size of this region compared to $\mu^{-1}$, the typical length scale of the relaxion, that determines whether a bubble is formed.
Let us then consider a simple model of the magnetosphere, in particular a rotating dipole (see e.g.~\cite{Garbrecht:2018akc}), in which
\beq
\vec{E} \cdot \vec{B} (r) = \left(\frac{\Bs^2 \Rs^6 \Os^2}{4 r^4}\right) \Theta(r-\Rs) \,.
\label{eq:dipole}
\eeq
Here $\Os$ is the angular velocity and $\Bs$ the magnetic field at the surface of the star, $r = \Rs$. We note that in general the $\vec{E} \cdot \vec{B}$ of a dipole rotating around the $z$ axis depends on the spherical coordinate $\theta$ as well as on the relative angle between the dipole and the axis of rotation of the star, $\alpha$. In \Eq{eq:dipole} we have simply integrated over $\theta \in [0,\pi]$ and taken $\alpha = \pi/4$.

The critical value of $\eta$ at which the minimum in vacuum ceases to be a minimum in the electromagnetic background is simply given by $\eta_c = \delta^2/(1-\delta^2)$. 
The value of $\delta^2$ at the $\ellm$-th shallow minimum is given, as in the QCD relaxion, by $\delta_{\ellm}^2 = \ellm \delta_{1}^2$, where $\delta_{\ellm = 1}^2$ is as in \Eq{eq:delta1qcd} with $\qcd \to \tc$.
The critical value of the electromagnetic field where $\eta = \eta_c$ is reached at a transition radius $\Rtem$ (equivalent to the radius of the dense star's core $\Rt$, see \Sec{sec:bubble}), given by
\beq
\Rtem = \left( \frac{g_{\phi \gamma \gamma} \Bs^2 \Rs^6 \Os^2}{4 \delta^2 \tc^4} \right)^{1/4} \, . 
\label{eq:rtem}
\eeq
Clearly, $\Rtem$ is much larger for shallow minima ($\delta^2 \ll 1$) than for deep ones ($\delta^2 \approx 1$), since the size of the critical electromagnetic field is much smaller for the former than for the latter.
The condition that $\Rtem > \Rs$, which is equivalent to the statement that the critical value of $\eta$ is reached somewhere before the surface of the star, is certainly necessary for a bubble to form (equivalent to the condition $\zeta(n_b) > \zeta_c$ in the relaxion models previously discussed). However, since $\vec{E} \cdot \vec{B}=0$ for $r < \Rs$ , the conditions for the formation and expansion to infinity of the bubble sets a lower bound on the size of the region $\Rtem - \Rs$ which is always more stringent than just $\eta(\Rs) > \eta_c$. For shallow minima, the condition for the formation of a $2 \pi f$ bubble is roughly given by
\beq
\Rtem - \Rs \gtrsim \frac{f}{\roll^2} \simeq \frac{f}{\tc^2} \, . \qquad (\textrm{technicolor; shallow})
\label{eq:shallowtc}
\eeq
This is the same condition leading to the escape of a bubble to infinity, since when $\delta^2 \ll 1$ the change in the potential from the inside to the outside of the transition region $r \sim \Rtem$ is barely appreciable. 
We can explicitly verify this is the case by considering the equation of motion of the bubble wall within the background electromagnetic field (see \cite{Balkin:2021zfd} for the equivalent in the case of a bubble wall within a star),
\beq
\sigma \ddot{R} = \epsilon - \frac{2\sigma}{R} - \sigma' \, , \qquad  \epsilon(R) = 2 \pi \roll^4 \left[ 1+ \frac{\delta^2}{1-\delta^2} \left(\frac{\Rtem}{R}\right)^4 \right] \, ,
\label{eq:eomem}
\eeq
where $\epsilon$, the energy density inside the bubble, changes with $R$ due to the fact that $\vec{E} \cdot \vec{B}$ and thus the effective rolling term $\roll^4(1+\eta)$ do.
It is then clear that for a shallow minimum, where $\delta^2 \ll 1$ and therefore $\sigma' = d\sigma/dR \simeq 0$, the condition for the bubble to escape is, to good approximation, given by \Eq{eq:shallowtc}.
Note that we have neglected $O(1)$ factors as we did in \Eqs{eq:condshallow}{eq:conddeep}, yet we expect them to be different here due to the non-spherical morphology of the system.

\begin{figure}[t]
	\centering
	\includegraphics[width=0.4\textwidth]{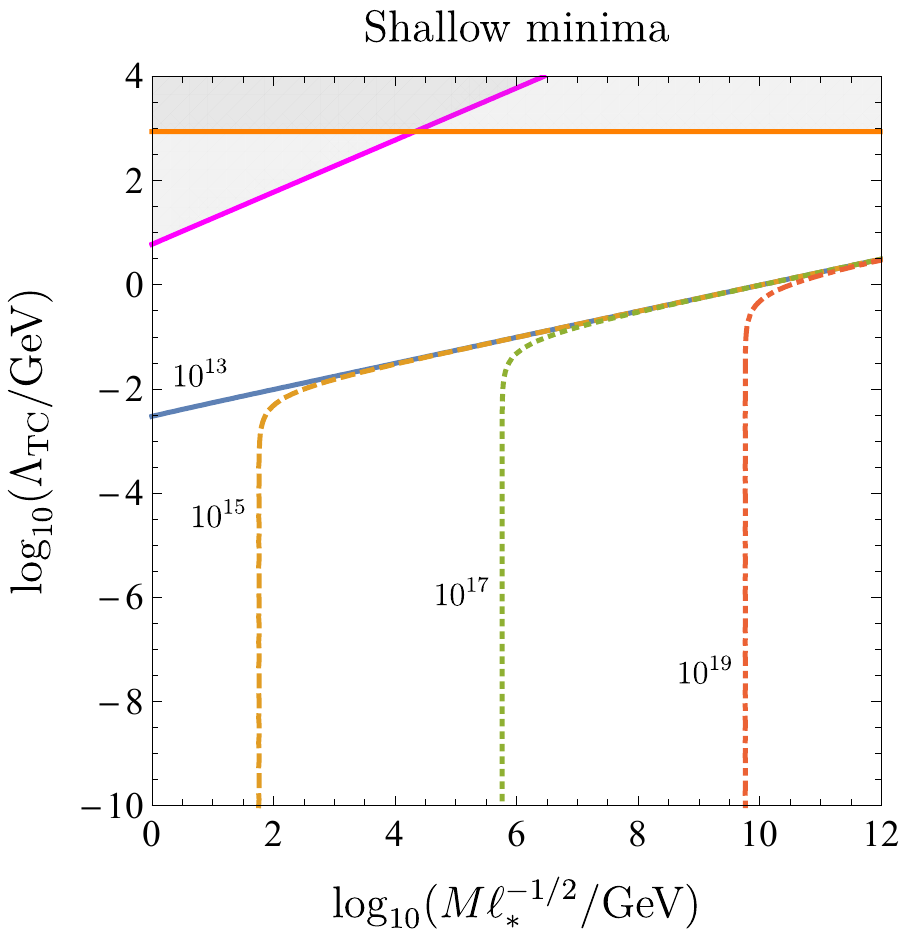}
	\hspace{1cm}
	\includegraphics[width=0.41\textwidth]{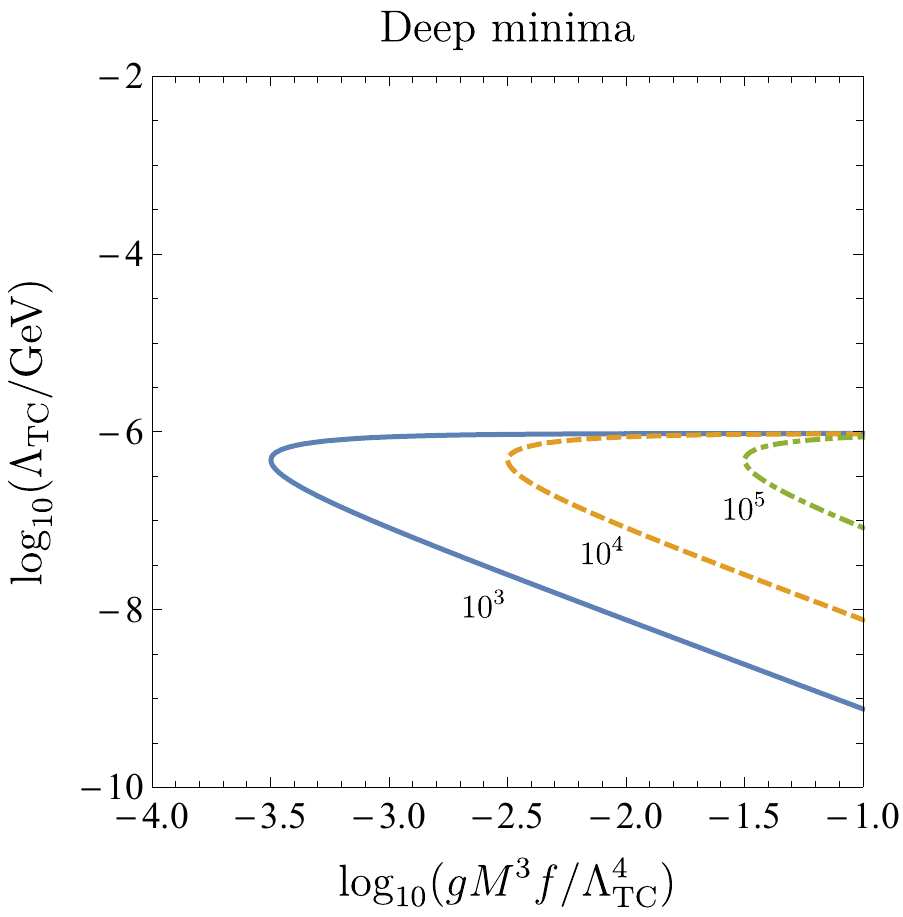}
	\caption{Regions excluded (below the labelled lines for several values of $f [\rm{GeV}]$) by the formation and escape of a TC relaxion bubble induced by the electromagnetic fields generated by rotating neutron stars. Left panel: for a bubble connecting shallow minima, $\delta^2 \ll 1$ or equivalently $\roll \simeq \tc$. For smaller values of the decay constant, the excluded regions coincide with the one for $f = 10^{13} \GeV$, the corresponding vertical boundary lying at smaller values of $M/\sqrt{\ellm}$. Right panel: for a bubble connecting deep minima, the depth parametrized by $gM^3f/\tc^4 = 1-\delta^2$. Note the different range for $\tc$ (recall for shallow minima $\tc \simeq \roll$) between the two plots.}
	\label{fig:boundstc}
\end{figure}

When the condition in \Eq{eq:shallowtc} is satisfied, the phase transition implies a change in vacuum energy that is experimentally too large for $2 \pi \roll^4 \gtrsim 10^2 \Lambda_0$, see \Eq{eq:lambdabound}. This allows us to exclude large regions of parameter space of the technicolored relaxion, as shown in the left panel of \Fig{fig:boundstc}.
To evaluate such a condition, we have taken as rotating neutron star properties, $\Rs = \Rns$ in \Eq{eq:refrs}, and typical values for the surface angular velocity and magnetic fields of neutron stars,
\beq
\Ons \approx 10 \, {\rm Hz}\, , \quad \Bns \approx 10^{10} \, {\rm T} \, ,
\label{eq:rot}
\eeq
see e.g.~\cite{Harding:2013ij}. The relaxion coupling to photons is given in \Eq{eq:tcem}, where we set $c = 1$ (and $\alpha \approx 1/137$); note that only if $c>0$ the rolling term is larger than in vacuum ($c<0$ would instead make the minimum deeper in the electromagnetic background).
In \Fig{fig:boundstc} (left panel), the region below a given labelled line is excluded, where each line corresponds to a different value of the relaxion decay constant (from $f = 10^{13}$ to $10^{19} \GeV$). For a fixed $f$, large values of $M/\sqrt{\ellm}$ are excluded irrespective of the value of the rolling term $\roll \simeq \tc$ as long as this is small. This is because the rolling term also controls the size of $\Rtem$, such that the condition \Eq{eq:shallowtc} becomes $\tc$-independent for large $M/\sqrt{\ellm}$. As $\tc$ increases, a certain critical value is reached where the size of the critical region quickly decreases and becomes smaller than $\Rns$. The condition $\Rtem \gtrsim \Rns$ is independent of $f$, which is why all the excluded regions share the same upper boundary.

Interestingly, the presence of large electromagnetic fields around neutron stars also leads to non-trivial constraints in the case of a bubble which connects deep minima.
If the condition $\eta > \eta_c$ is satisfied, the condition for formation of a deep bubble is easily satisfied. This is because since $\eta_c \gg 1$, the slope of the potential is much larger in the region $r \lesssim \Rtem$ than in vacuum.
The relevant condition that leads to a phase transition is then the one concerning the escape of the bubble. 
We derive such a condition under the conservative simplification that past the transition radius the relaxion potential approximately returns to its in-vacuo form, i.e.~$\eta(r > \Rtem) = 0$. This is justified by the rapid decrease of $\vec{E} \cdot \vec{B}$, and therefore of $\eta$, with $R$, see \Eq{eq:dipole}. Then, our escape condition follows from requiring that $\ddot{R} > 0$ in \Eq{eq:eomem}, taking the minimal value of $\epsilon$, that is $2 \pi \roll^4$, and the value of the tension force at the transition radius, that is $2 \sigma/\Rtem$,
where note that due to our simplification we have $\sigma'(r > \Rtem) = 0$. Then, a deep bubble escapes to infinity if
\beq
\Rtem - \Rs \gtrsim \frac{f}{\roll^2} \frac{1}{\sqrt{1-\delta^2}} \simeq \frac{\tc^2}{g M^3} \, , \qquad (\textrm{technicolor; deep})
\label{eq:deeptc}
\eeq
where we recall that $1-\delta^2 = \roll^4/\tc^4 \ll 1$ and $\roll^4 = g M^3 f$. As in the case of relaxion bubbles seeded by baryon density, this condition implies that it is more difficult for a deep bubble to escape than a shallow one. The underlying reason is the same as well, in vacuum the bubble-wall tension is dominated by large potential barriers.
Nevertheless, as shown in the right panel of \Fig{fig:boundstc}, for not too small values of $\roll^4/\tc^4$, a phase transition can still be induced between deep minima of the technicolored relaxion. As expected, the excluded regions (to the right of a given line) correspond to small values of the relaxion decay constant (up to $f \sim 10^{5} \GeV$) and of the technicolor confinement scale. We further note that when $\tc$ gets small for fixed $\delta$, and even though $\Rtem$ gets larger, the gain in volume energy is too little to push the bubble outwards, preventing the phase transition from expanding to the entire universe. 
Finally, we note that the lower bounds we derived on $\tc$ can be rephrased, given the collider constraints on $v'$, as a lower bound on $m_U$ in \Eq{eq:tcscale}.

%%%%%%%%%%%%%%%%%%%%%%%%%%%%%%%%%%%%%%%%%%%
\subsection{Dark compact objects} \label{sec:dark}

In this section we entertain the possibility that there are dark compact objects~\cite{Foot:1999hm,Narain:2006kx,Spolyar:2007qv,Sandin:2008db,Kouvaris:2015rea,Gresham:2018rqo,Curtin:2019ngc,Curtin:2019lhm,Hippert:2021fch} in the universe. These dark stars, similar to standard stars, can induce the instability of a metastable vacuum.
This would be particularly relevant when the periodic term in the landscape potential \Eq{eq:Vgeneric} arises from dark dynamics, as in the case of the non-QCD relaxion (especially if the scalar is sitting in a deep minimum, as in e.g.~\cite{Espinosa:2015eda}) or in models where the barriers are Higgs independent \cite{Hook:2016mqo,Fonseca:2018xzp}. In addition, we show that this possibility opens the door to late-time phase transitions for which the associated change in vacuum energy is a priori experimentally allowed. As shown in \Eq{eq:lambdachange}, in the case of standard stars this can only happen for quite low values of $f$ and if the bubbles are seeded solely by the largest stars known to date, with $\Rs \sim 10^3 R_\odot$.%
\footnote{Let us point out as well that phase transitions for which $|\Delta \Lambda| \lesssim \Lambda_0$ could have implications for our understanding of the cosmological constant problem \cite{Balkin:2021zfd}.}

Let us assume then a new species of fermion, which we refer to as dark baryon, whose relic abundance is non-negligible and which constitutes the main component of the dark stars (yet not necessarily making up all of the dark matter). Let us note right away that the existence of these stars requires non-trivial dynamics by which the dark fermion can dissipate their kinetic energy, accumulate and eventually form a compact object. If this is the case, the smaller dark stars will only be sustained by the Fermi degeneracy pressure associated with the dark baryon, thus with typical radii and densities
\beq
\Rnsd \sim \sqrt{8 \pi} \frac{\mplanck}{\mdark^2} \, , \quad \ndark \sim \mdark^3 \, ,
\label{eq:rsdark}
\eeq
where $\mdark$ is the mass of the dark baryon. 

Before moving on, we note that such a dark baryon is in fact motivated by the non-QCD relaxion, whose simplest UV realization crucially involves $N_{\tilde f}$ flavours of SM-neutral fermions, $N$, charged under a new confining $SU(N_{\tilde{\text{\tiny C}}})$ gauge group.
The associated IR scale, which controls the size of the relaxion barriers, is given by $\noqcd^4 \simeq 4 \pi f_{\tilde \pi}^3 \mN$ where $\mN = \mN (h^2) \ll 4 \pi f_{\tilde \pi}$ is the mass of the dark quarks (taken degenerate for simplicity), whose dominant contribution is proportional to the square of the Higgs VEV, \Eq{eq:higgsvev}.
The mass of the dark baryons, analogous to the QCD baryons, receives two contributions,
\beq
\mdark = \tilde{m}_0 + \tilde{\sigma}(\mN) \, ,
\label{eq:mbdark}
\eeq
where $\tilde{m}_0$ is purely due to the dark strong dynamics while $\tilde{\sigma}$ is the analogue of the pion-nucleon sigma term of QCD. Likewise, at finite dark density, the barriers decrease according to $\noqcd^4 \to \noqcd^4 (1-\zeta)$, where in the linear approximation,
\beq
\zeta(\ndark) \simeq \frac{\tilde{\sigma} \ndark}{\noqcd^4} \sim \frac{\mdark^3}{4 \pi f_{\tilde \pi}^3} \, .
\label{eq:zetadark}
\eeq
where in the last equality we have used \Eq{eq:rsdark} and the fact that $\tilde{\sigma} \simeq a \mN$ if $\mN \ll f_{\tilde \pi}$, where $a = \mathcal{O}(1)$ (yet note that in QCD the analogous coefficient is rather $\approx 10$).
Therefore, for a sufficiently large dark baryon mass, yet small enough to retain perturbative control, densities can be enough to seed the formation of a bubble.
Finally, the condition that the system is large enough for the bubble to escape, assuming for simplicity that $\delta \sim 1$, and given that the size of the dark neutron star is controlled by $\mdark$, yields the condition
\beq
\mdark \lesssim \noqcd \sqrt{\frac{\mplanck}{f}} \, .
\eeq
Interestingly, the change in vacuum energy \Eq{eq:lambda} associated with such a relaxion bubble is controlled by $\mdark$ as well. Considering again for simplicity the case $\roll \sim \noqcd$,
\beq
- \Delta \Lambda \gtrsim \mdark^4 \left( \frac{f}{\mplanck} \right)^2 \approx 6 \times 10^{-3} \, \Lambda_0 \left( \frac{\mdark}{10 \, {\rm keV}} \right)^4 \left( \frac{f}{10 \TeV} \right)^2 \,,
\label{eq:lambdadark}
\eeq
where the values of the dark mass and decay constant have been taken to illustrate that the change can be small enough as to avoid any trivial experimental inconsistency between the early and late universe.
This gives rise to the exciting possibility that the change in the relaxion minimum could be detected with future cosmological measurements. In addition, if $\mdark$ or $f$ are small enough and the dark stars are dense and large enough to destabilize many relaxion minima (i.e.~$N \gg 1$, see \Eq{eq:N} and \App{sec:Nlarge}), the effects of the continued phase transitions originating from the ongoing creation of relaxion bubbles interpolating between lower and lower pairs of consecutive minima
could resemble the time evolution of a quintessence field as dark energy \cite{Garriga:2000cv}.

Finally, we note that the in-vacuo relaxion mass, for the range of relaxion parameters where the change in vacuum energy is smaller than its current value, is
\beq
m_\phi \lesssim \frac{1}{f} \sqrt{\frac{\Lambda_0}{2 \pi}} \approx 2 \times 10^{-16} \, {\rm meV} \left( \frac{10 \TeV}{f}\right) \, ,
\label{eq:massdark}
\eeq
which is, as expected, extremely small. Accordingly, the size of the dark compact object, $\Rnsd \sim 1/m_\phi$, is very large
\beq
\Rnsd \gtrsim f \sqrt{\frac{2 \pi}{\Lambda_0}} \approx 1 \times 10^{9} \km \left( \frac{f}{10 \TeV}\right) \, ,
\label{eq:stardark}
\eeq
which for this value of $f$ is roughly the size of the solar system.

%%%%%%%%%%%%%%%%%%%%%%%%%%%%%%%%%%%%%%%%%%
%%%%%%%%%%%%%%%%%%%%%%%%%%%%%%%%%%%%%%%%%%
\section{Conclusions} \label{sec:conclusions}

The relaxion mechanism provides a solution to the electroweak hierarchy problem by postulating a landscape of vacua where the potential barriers between minima depend on the Higgs VEV, enabling vacuum selection with a small electroweak scale via a period of dynamical evolution. 
An immediate concern of postulating a multi-vacuum potential is whether the selected vacuum is stable on cosmological scales.
While one usually considers vacuum transitions of quantum or finite temperature origin, in this work we focused on the certainly less studied case of phase transition seeded by finite density objects such as stars. 
This type of transition, if it occurs, would take place at a much later period in the cosmic history compared to e.g.~temperature-driven phase transitions. 
Such late-time phase transitions are constrained by cosmological measurements and could even have lethal implications for our universe.
As a result, we were able to place new bounds on various relaxion models, ruling out regions of parameter space where such forbidden phase transitions would take place once stars are formed. 
To this end, we relied on the formalism and results of~\cite{Balkin:2021zfd}, where density-driven phase transitions were studied in detail, and adapted them to a typical relaxion potential.

We showed that the connection of the relaxion with the Higgs is precisely what is behind the sensitivity of the relaxion vacua to finite density effects.
In particular, in realizations where the potential barriers are generated by QCD dynamics, baryonic densities decrease the chiral symmetry breaking scale, leading to the possibility of QCD-relaxion bubbles. In realizations where instead new confining dynamics is responsible for the barriers, the change in the Higgs VEV due to the background nucleons and muons, although small, is sufficient in some regions of parameter space to induce the formation of bubbles. 
Generically, we found that these bubbles easily escape from the stars where they are formed: neutron stars, white dwarfs, or main-sequence stars, depending on how small the overall scale of the non-QCD relaxion potential is.
Once the bubble escapes, the associated change in the vacuum energy of the universe is too large to conform with early versus late cosmological measurements of the energy budget of our universe.
Therefore, we set new bounds on relaxion models, ruling out those regions of parameter space where expanding relaxion bubbles could have been generated during star formation (at redshifts $z \sim 10$).

Notably, we discovered that not only matter density but an electromagnetic background can destabilize a metastable vacuum.
This possibility is motivated by some constrained yet still viable realizations of the relaxion, those in which the scalar field has large couplings to photons.
We found that the large electric and magnetic fields of magnetars/pulsars destabilize the metastable minimum and lead to a phase transition that cannot be confined.
Moreover, in this scenario, the transition can occur not only for shallow minima, but also for metastable vacua in which there is a hierarchical separation between the energy difference and the potential barrier between the minima.

In general, relaxion phase transitions leading to a very small change in vacuum energy compared to its measured value could in fact have been induced by the formation of large dense objects in the universe. This is the case for very low relaxion decay constants and for the largest stars in the universe acting as seeds. 
We also considered the possibility that these naively harmless phase transitions may be the result of the formation of very large dark stars. Such stars would be sustained by the Fermi degeneracy pressure associated with the light stable dark baryons motivated by the non-QCD relaxion.

Finally, the new type of bound derived in this work for the relaxion landscape, namely vacuum instability induced by dense objects, could be relevant for other landscapes if subject to finite density deformations. These deformations are generically expected if the vacua are tied to the electroweak scale.

%%%%%%%%%%%%%%%%%%%%%%%%%%%%%%%%%%%%%%%%%%

\section*{Acknowledgments}

We would like to thank Geraldine Servant for useful discussions. The work of RB, JS, KS, SS and AW has been partially supported the Collaborative Research Center SFB1258, the Munich Institute for Astro- and Particle Physics (MIAPP), and by the Excellence Cluster ORIGINS, which is funded by the Deutsche Forschungsgemeinschaft (DFG, German Research Foundation) under Germany's Excellence Strategy -- EXC-2094-390783311.
RB is additionally supported by grants from NSF-BSF, ISF and the Azrieli foundation.

%%%%%%%%%%%%%%%%%%%%%%%%%%%%%%%%%%%%%%%%%%
%%%%%%%%%%%%%%%%%%%%%%%%%%%%%%%%%%%%%%%%%%

\newpage

\appendix

%%%%%%%%%%%%%%%%%%%%%%%%%%%%%%%%%%%%%%%%%%
\section{Formation and escape of $N\gg1$ bubbles}\label{sec:Nlarge}

In the main text we have concentrated on bubbles interpolating between two consecutive relaxion minima, located at the period $\ellm$ outside the bubble and at $\ellm + 1 $ inside it.
However, the relaxion displacement at the core of the star could be much larger than $2 \pi f$, in particular $\Delta \phi(0) \sim 2\pi f N$ with $N \gg 1$ is expected to naturally occur for relaxion bubbles at densities significantly above the critical one, see \Eq{eq:N}. In this appendix we provide a discussion of the fate of the relaxion bubbles in such a situation, following closely \cite{Balkin:2021zfd} while avoiding the detailed derivation presented there.

For the first in-density minimum to be reached, the energy in the in the field's gradient, $\sim (2 \pi f N/\Rt)^2$, needs to be compensated by the gain in potential energy inside the bubble, $\sim 2 \pi \roll^4 N$. This can only happen if the core of the star is large enough
\begin{equation}
\Rt \gtrsim \frac{\sqrt{N}}{\mroll} \,,
\label{eq:rtlarge}
\end{equation}
where we recall that $\mu = \roll^2/f$. If this is the case, a large bubble with a field displacement of $\Delta \phi(0) \sim 2\pi f N$ is fully formed.
The properties of such a bubble, that is its volume energy density and tension can be simply estimated as,
\begin{align}
\epsilon & \sim \mroll^2 f \Delta \phi(0) \sim 2 \pi \roll^4 N \,, \\ 
\sigma & \sim\Delta \phi(0) \sqrt{\epsilon} \sim \roll^2 f (2 \pi  N)^{3/2} \,.
\end{align}

\begin{figure}[t]
	\centering
	\includegraphics[width=0.6\textwidth]{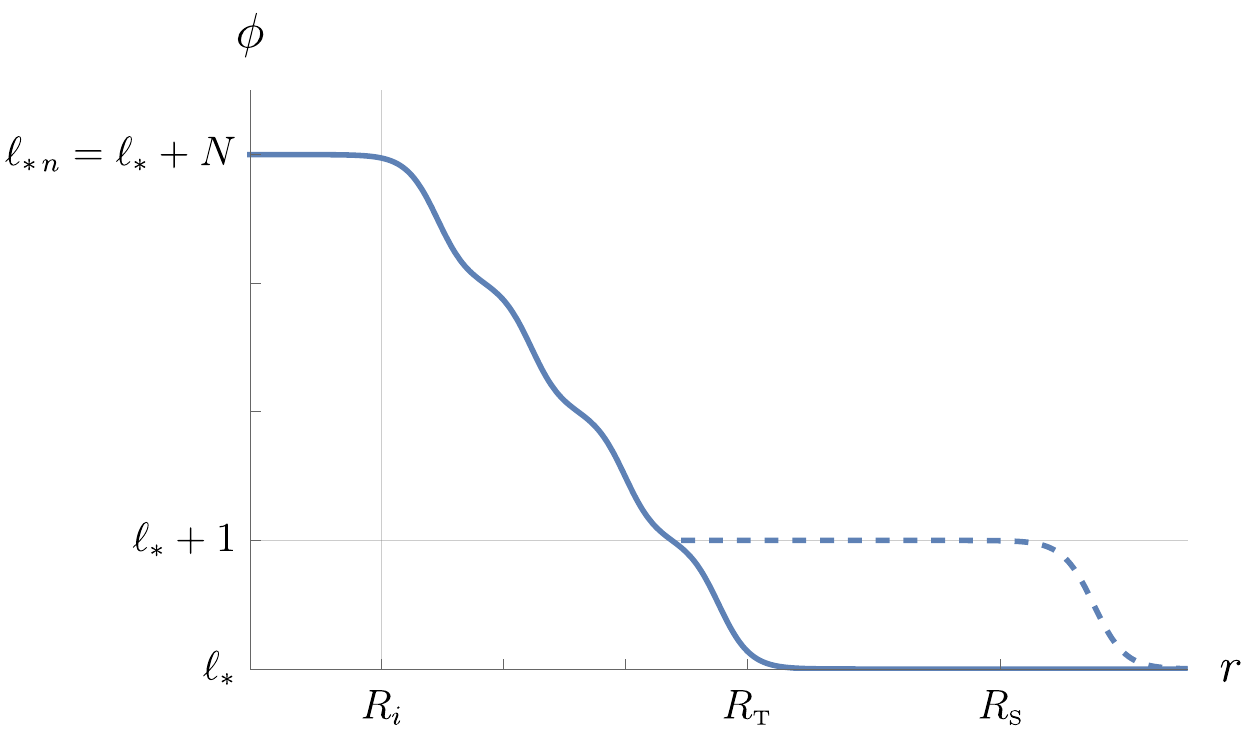}
	\caption{$N \gg 1$ bubble in equilibrium at $R \sim \Rt$ (solid) and sub-bubble that has already escaped from the star (dashed).}
	\label{fig:subbubble}
\end{figure}

This large relaxion bubble can be thought as made of a series of $N$ sub-bubbles, each corresponding to a field displacement w.r.t.~to the next of $2 \pi f$, see \Fig{fig:subbubble}. This is motivated by the fact that in vacuum each relaxion minimum is separated by a potential barrier, so the escape of relaxion bubbles from the star takes place in discrete steps, starting with the outermost bubble, within which the relaxion sits just one period away, i.e.~at $\ellm + 1$, from its in-vacuo value.
This sub-bubble has then a volume energy density and tension
\begin{align}
\epsilon_{\rm{sub}} & \sim 2\pi\roll^4 \,, \\ 
\sigma_{\rm{sub}}(\Rt) & \sim \roll^2 f (2\pi N)^{1/2} \,.
\end{align} 
Note in particular that the wall tension is enhanced by a factor $\sqrt{N}$ w.r.t.~the one of a standard relaxion bubble.
It is this enhancement that facilitates the sub-bubble escape from the star. Indeed, the characteristic contracting force of bubbles propagating through a star, $\sigma' \sim \Delta \sigma/\DRt$ associated with a radius-dependent tension, is mitigated because $\Delta \sigma \sim \sigma_{\rm{sub}}(\Rs)-\sigma_{\rm{sub}}(\Rt)$ decreases (or could even become negative) at large $N$.
We have explicitly verified this effect via numerical simulations of the sub-bubble's dynamics.

The condition for the expansion, beyond the confines of the star and towards infinity, of this $2 \pi f$ bubble then coincides with the condition that the bubble does not contract in vacuum, $\epsilon_{\rm{sub}} \gtrsim 2 \sigma_{\rm{sub}}(\Rs)/\Rs$, where $\sigma_{\rm{sub}}(\Rs)$ is the tension of the wall in vacuum.
For shallow minima, since the potential barrier \Eq{eq:barrier} is very small, this tension is simply $\sigma_{\rm{sub}}(\Rs) \sim 2 \pi f \sqrt{\epsilon_{\rm{sub}}}$. Therefore, the condition reads
\beq
\Rs \gtrsim \frac{f}{\roll^2}\,, \qquad (\textrm{sub-bubble; shallow})
\label{eq:Rsminsub}
\eeq
which is automatically satisfied given that $\Rs > \Rt$ and \Eq{eq:rtlarge}.
Instead, for deep minima we can use the thin-wall approximation to compute the wall's tension $\sigma_{\rm{sub}}(\Rs) \simeq 8  f \br^2$, and the correct condition is then
\beq
\Rs \gtrsim \frac{f}{\roll^2} \frac{1}{\sqrt{1-\delta^2}} \,. \qquad (\textrm{sub-bubble; deep})
\label{eq:condsub}
\eeq

Finally, let us note that if the conditions for a sub-bubble to escape are satisfied, this will also be the case for subsequent sub-bubbles (with $N \to N-1$), up until $N$ becomes so small that the enhancement in the sub-bubble tension $\sigma_{\rm{sub}}$ is not enough to guarantee that $\epsilon \gtrsim \sigma'$, that is that the outwards pressure due to the gain in ground state energy is not enough to overcome the tension's gradient.

%%%%%%%%%%%%%%%%%%%%%%%%%%%%%%%%%%%%%%%%%%
\section{Opposite-sign back-reaction} \label{sec:opposite}

The discussion in the main text has been restricted to densities that, while allowing for the field to classically move inside the star, i.e.~$\zeta(n) > \zeta_c = \delta^2$, are still such that the back-reaction term in the potential is non-vanishing, even if negligibly small. In this section we want to consider instead the possibility that $\zeta(n) > 1$, such that the wiggles not only vanish but change sign in some region inside the star.

In this situation, we can identify another inner core radius $\Rt'$, such that for $r < \Rt'$, minima of the in-density potential reappear due to opposite-sign barriers. This is fixed by $\zeta(\Rt') = 2 - \delta^2$, where $\delta$ is defined, as in vacuum, as the difference between the size of the rolling and back-reaction terms at the minimum of interest, see \Eq{eq:delta}.
Because of the different sign of the back-reaction term, the minima are now shifted by $\pi$ with respect to those for $\zeta < 1$, i.e.~they are located at $( 2\pi \ell_{* \, n} + \theta_{*\,n} + \pi) f$.
The field displacement, or equivalently the value of $\ell_{* \, n}$ where the relaxion sits for $r < \Rt'$, is determined by the size of the region between the two core radii, $\DRt' \equiv \Rt - \Rt'$.
This difference sets, in the same fashion as \Eq{eq:rtlarge}, the field displacement from the in-vacuo value, 
$\Delta \phi(\Rt')/f \sim (\mu \DRt')^2$.
From this point on, the fate of the bubble (or sub-bubble) is not much different than what already discussed in \Sec{sec:bubble} and \App{sec:Nlarge}. In particular, if $\DRt' \gg \mu^{-1}$, the relevant dynamics is that of a $2 \pi f$ sub-bubble, for which the condition \Eq{eq:Rsminsub} determines if it escapes the star and expands to infinity. The only subtlety arises for $\Delta \phi(\Rt') = \pi f$. In this case the potential energy density of the bubble receives a contribution from the back-reaction term in addition to the rolling term. We find $\epsilon \simeq \pi \roll^4 + 2 (\zeta-1) \br^4$, where we recall that we are dealing with densities such that $\zeta > 1$. While for sufficiently large $\zeta$ this extra contribution naively helps the bubble expand, as soon as the bubble wall goes through the outer transition region of the star, $r > \Rt$, $\epsilon$ decreases because the relaxion value inside the bubble, $(2 \pi \ellm + \pi)f$, is not a minimum of the potential in vacuum. This eventually prevents the bubble from escaping.
This type of confined bubble (yet with $\roll = 0$) has been found to be a plausible consequence of the QCD axion \cite{Balkin:2020dsr}, or special deformations thereof \cite{Hook:2017psm}, in neutron stars.

%%%%%%%%%%%%%%%%%%%%%%%%%%%%%%%%%%%%%%%%%%
\section{Bubble interactions with the environment} \label{app:environment}

In the main text we have treated the density profile as a non-dynamical classical background field, upon which a non-trivial scalar field develops. In this appendix we study in some more detail the interactions of the scalar bubble with the dense environment. 

Let us discuss first the force exerted by individual nucleons, $N$, on the bubble wall. We focus on the case of the QCD relaxion, since the interactions of non-QCD relaxions with protons and neutrons are much weaker, being mediated by Higgs exchange.
The interaction with nucleons is of the form $\sim \sigma_{\pi N} \bar{N}N \cos(\phi/f)$. This constitutes a contribution to their mass that depends on the relaxion field, and therefore on space-time, $m_{\pi N}(r) \sim \sigma_{\pi N} \cos(\phi(r)/f)$.
Recall that most of the mass of a nucleon comes from a term independent of the quark masses and thus independent of $\phi$, $m_N = M_{B} + m_{\pi N}$ with $M_{B} \gg m_{\pi N}$.
It is precisely this interaction of the relaxion with nucleons that gives rise to the leading linear correction to the back-reaction term in the limit of small baryonic densities, after substituting $\bar{N}N \to n_b$, as given in \Eq{eq:zetaHF}.
Note however that for this treatment to hold, one is implicitly assuming that the scalar field interacts classically with the density profile, or in other words that single nucleons are able to penetrate the bubble wall with negligible quantum-mechanical reflection.

Let us have a look then at the one-dimensional quantum mechanics of a nucleon in the potential associated with its space-time dependent mass, $m_{\pi N}(r)$.
This follows the discussion in \cite{Hook:2019pbh}, with some important modifications.
Before proceeding, let us note there are two qualitatively different types of potentials depending on the relaxion profile. 
For bubbles in which the scalar field displacement is $\Delta \phi(0) = 2\pi f$ (our focus in the main text) the overall change of the nucleon mass between inside and outside the bubble vanishes, while at the center of the bubble wall, where $\Delta \phi = \pi/2$, $m_{\pi N} \sim - \sigma_{\pi N}$; the potential thus resembles a well. If instead the field displacement is $\Delta \phi(0) = \pi f$ (briefly discussed in \App{sec:opposite}), we have that the change from outside to inside the bubble $\Delta m_{N} \sim - 2 \sigma_{\pi N}$, and the potential is a downwards step. 
In any of these cases, in order to properly compute the force that the nucleons exert on the bubble wall, it is important to realize that the relevant scales of the problem are the thickness of the wall, of order $\mu^{-1} \sim f/\roll^2$, and the nucleon wavelength, given by $\lambda_N \sim 1/m_N v$, where $v$ is the relative velocity of the nucleons with respect to the wall, which we expect to be non-negligible (either because of their Fermi momentum, temperature, or the initial yet small velocity that the bubble acquires when it forms). We therefore expect $\lambda_N \mu \ll 1$, which already indicates that the nucleons interact with the potential classically. Focussing for concreteness on the step-like potential, we can go further and split it in $J$ small patches, each of them of size $\lambda_N$, where quantum mechanical effects become important. In each step the nucleon mass decreases by an amount $\delta m_N = |\Delta m_{N}|/J \sim 2 \sigma_{\pi N}/J$.
Therefore, for each step the quantum-mechanical reflection coefficient is given by
\beq
\mathbb{R}_1 = \frac{(k-k')^2}{(k+k')^2} \simeq \frac{\delta m_N^2}{4 m_N^2 v^4} \,,
\eeq
where $k$ is the momentum of the nucleon before traversing the wall, and $k'$ its momentum once inside the bubble. In the second equality we have taken the non-relativistic limit, $k \simeq m v$ and $k' \simeq \sqrt{k^2 + 2 m_N \delta m_N}$, and expanded in $\delta m_N$.
Taking into account all the $J$ barriers, the total reflection coefficient is bounded by
\beq
\mathbb{R} \lesssim J  \mathbb{R}_1 \simeq \frac{\Delta m_{N}^2}{4 m_N^2 v^4 J} \, ,
\eeq
which vanishes in the limit $J \gg 1$. 
This agrees then with the naive expectation that for $\lambda_N \mu \ll 1$ the system behaves classically, without reflection.
In fact, one can also compute the total force on the bubble wall associated with the gain in momentum of the nucleons as they pass through it.
For a single nucleon, after going down all the $J$ steps, the force is $f_N = J (k' - k) \mathbb{T} \simeq |\Delta m_{N}|/v$. Therefore, for an ensemble of nucleons with density $n_b$, the total force reads
\beq
F_N = n_b v f_N \sim 2 \sigma_{\pi N} n_b \, .
\eeq
This precisely matches the piece of the volume force associated with the change of the potential barriers derived in \App{sec:opposite} (opposite-sign back-reaction), $\sim 2 \zeta \br^4 \sim 2 \sigma_{\pi N} n_b$.
For the potential-well case (i.e.~for $2 \pi f$ bubbles), with this quantum-mechanical treatment we find, as expected, $F_N = 0$. However, given that the wall appears to the individual nucleons as a classical potential well, these tend to accumulate at the wall, i.e.~the density (as well as temperature) increases at the wall; this in turn means that the wall gets thicker.
Therefore, it appears that as the bubble expands through the star, it carries with it a local (of size $\mu^{-1}$) increase in density. However, for most of the star the density profile remains unaltered.

\bibliography{bib/finiteDensityRelaxion}
\bibliographystyle{jhep}

\end{document}